\documentclass[a4paper,fleqn,usenatbib,useAMS]{mnras}

\usepackage{mathptmx}

\usepackage[T1]{fontenc}
\usepackage{ae,aecompl}

\usepackage{graphicx}	
\usepackage{amsmath}	
\usepackage{amssymb}	
\usepackage{newtxtext,newtxmath}

\newcommand{\logg}{\log\,g}
\newcommand{\teff}{T_{\rm eff}}
\newcommand{\bz}{\langle B_z \rangle}

\newcommand{\kms}{km\,s$^{-1}$}

\title[Magnetic field geometry of NGC\,1624-2]{
Magnetic field geometry and magnetospheric environment of the strongly 
magnetic Of?p star NGC\,1624-2
}
\author[S. P. J\"arvinen et al.]{
S.~P.~J\"arvinen$^{1}$\thanks{E-mail: sjarvinen@aip.de},
S.~Hubrig$^{1}$,
M.~Sch\"oller$^{2},$
M.~K\"uker$^{1}$,
I.~Ilyin$^{1}$,
S.~D.~Chojnowski$^{3}$
\\
$^{1}$Leibniz-Institut f\"ur Astrophysik Potsdam (AIP), An der Sternwarte~16, 14482~Potsdam, Germany\\
$^{2}$European Southern Observatory, Karl-Schwarzschild-Str.~2, 85748 Garching, Germany\\
$^{3}$Apache Point Observatory and New Mexico State University, P.O. Box 59, Sunspot, NM, 88349-0059, USA
}

\date{Accepted XXX. Received YYY; in original form ZZZ}

\pubyear{2020}

\begin{document}
\label{firstpage}
\pagerange{\pageref{firstpage}--\pageref{lastpage}}
\maketitle

\begin{abstract}
\mbox{NGC\,1624-2} is an O7f?p star with a reported probable polar magnetic 
field strength $\ge$ 20\,kG, which is the strongest magnetic field ever 
measured in an O-type star. We study the variability of the mean longitudinal 
magnetic field $\bz$ and the mean field modulus $\langle B \rangle$ to obtain 
constraints on its field geometry. Only one magnetic pole is observable over 
the rotation cycle. The approximately sinusoidal variation of $\bz$ and the 
ratio of the values of the extrema of $\langle B \rangle$ indicate that there 
is an important component of the field that is dipolar. The $\bz$ values 
measured over the rotation cycle are in the range from $-0.2$ to 4.5\,kG, 
whereas the values for $\langle B \rangle$ vary between 9 and 12\,kG. The 
$\bz$ values obtained using the \ion{O}{iii} $\lambda$7455 emission line are 
in the range from 0.4 to 2.3\,kG and show a variability pattern similar to 
that detected for the absorption lines. The fact that the phase of the $\bz$ 
minimum coincides with the phase of the $\langle B \rangle$ maximum, 
indicates that the field structure must significantly depart from a centred 
dipole. Further, we discuss the nature of the observed variable Stokes~$V$ 
profiles corresponding to a longitudinal field of negative polarity detected 
in the emission \ion{He}{i} lines and present the first MHD numerical 
simulations of the gas flow in the magnetosphere of this star.
\end{abstract}




\section{Introduction}
\label{sec:intro}

Recent observations indicate that probably only a small fraction of about
$7\pm3$\% of O-type stars with masses exceeding 18~${\rm M}_\odot$ 
\citep{Grunhut2017} 
and about $6\pm3$\% of early B- and O-type stars 
\citep{Schoeller2017}
have measurable, mostly dipolar magnetic fields. Among the polarimetrically 
studied O-type stars, all five Galactic stars with Of?p classification 
possess measurable magnetic fields. These stars exhibit recurrent, and 
apparently periodic, spectral variations in Balmer, \ion{He}{i}, 
\ion{C}{iii}, and \ion{Si}{iii} lines, sharp emission or P\,Cygni profiles in 
\ion{He}{i} and the Balmer lines, and a strong \ion{C}{iii} blend in emission 
around 4650\,\AA{} 
\citep{Walborn1972}. 
The presence of variable emission in the \ion{C}{iii} blend and in H$\alpha$ 
in Of?p stars is indicative of circumstellar structure, related to their 
magnetospheres. Previous measurements of the mean longitudinal magnetic field 
for the O7f?p star \mbox{NGC\,1624-2} carried out with ESPaDOnS (Echelle 
SpectroPolarimetric Device for the Observation of Stars) at the
Canada-France-Hawaii Telescope between 2012 February 1 and 9 indicated
$\bz =5.35\pm0.5$\,kG with a polar magnetic field strength of about 20\,kG, 
which is the strongest magnetic field ever measured in an O-type star
\citep{Wade2012a}. 
All the other Of?p stars with studied magnetic field geometries have 
$B_{\rm pole}\le2.6$\,kG 
\citep{David-Uraz2019}. 
The existence of a giant wind-fed dynamical magnetosphere around 
\mbox{NGC\,1624-2} was discussed by 
\citet{Petit2015}. 
A very recent study by
\citet{Kurtz}
reported the detection of coherent pulsation modes in \mbox{NGC\,1624-2}
based on Transiting Exoplanet Survey Satellite 
\citep[{\em TESS};][]{tess}
high-cadence photometry.

The study of spectral variability using archival optical spectroscopic 
observations indicated $v\,\sin\,i \le3$\,\kms{} and a rotation period of 
$157.99\pm0.94$\,d 
\citep{Wade2012a}. 
Due to the presence of the strong magnetic field, slow rotation, and the low 
projected rotational velocity, a few spectral lines showed magnetic splitting 
into Zeeman components corresponding to a maximum mean magnetic field modulus 
of about $\langle B \rangle =14\pm1$\,kG. Furthermore, 
\citet{Wade2012a} 
reported the detection of reversed Stokes~$V$ profiles associated with weak, 
high-excitation emission \ion{O}{iii} lines, and suggested that they may form 
in the low-velocity plasma confined in closed magnetic loops above the 
stellar surface. They measured a longitudinal field $\bz =2.58\pm0.7$\,kG, 
which is by more than a factor of 2 smaller than the field measured using 
photospheric absorption lines. On the other hand, the emission lines 
\ion{He}{i}~5876, 6678, and 7065 exhibited Stokes~$V$ profiles with signs 
consistent with those of the photospheric absorption lines, but yielding a 
negative mean longitudinal magnetic field. Unfortunately, no field 
measurements using these lines were presented in the work of 
\citet{Wade2012a}.
 
The \ion{He}{i} emission lines are marked in the paper of 
\citeauthor{Wade2012a} 
as ``wind lines''. This expression was already used by 
\citet{Howarth}
for H$\alpha$, where the authors discuss some lines as contaminated by 
``windy'' emission. Several studies of magnetic O-type stars concluded that 
these lines are not formed in the wind, but are possibly resulting from 
confined circumstellar material
\citep[e.g.][]{Grunhut2009}.
In the context of the magnetically confined wind-shock model, it was 
suggested that most emission comes from scattering by the cooling disk and 
very little from the wind 
\citep{Donati2002}.

The detected remarkable characteristics of \mbox{NGC\,1624-2} make it an 
excellent laboratory to study the impact of a very strong magnetic field on 
the behaviour of different elements in the stellar photosphere and 
circumstellar environment. In this work, we discuss for the first time the 
magnetic field geometry of \mbox{NGC\,1624-2} based on the variability of the 
longitudinal magnetic field measured using photospheric lines, the 
longitudinal field measurements using the emission \ion{O}{iii} and 
\ion{He}{i} lines, and the variability of the magnetic field modulus over the 
rotation period. Finally, using the {\sc NIRVANA} code, we discuss the first 
magnetohydrodynamical (MHD) numerical simulations carried out to understand 
the field geometry, mass distribution, and the gas motions in the vicinity of 
\mbox{NGC\,1624-2}.


\section{Observations}
\label{sect:obs}

A few polarimetric observations were obtained at the beginning of 2012 
\citep{Wade2012a}
with the ESPaDOnS spectropolarimeter installed at the 3.6\,m 
Canada-France-Hawaii Telescope (CFHT) together with one observation from its 
twin Narval, installed at the 2\,m Bernard Lyot Telescope (BLT) on Pic-du-Midi.
Six additional ESPaDOnS observations from 2012 September to 2013 January were
presented by 
\citet{Grunhut2017}.
Subsequently, sixteen ESPaDOnS observations were obtained between 2013 August 
and 2015 September. They are described in more detail in
\citet{daviduraz2020}.
All the ESPaDOnS data are publicly available in the CFHT Science 
Archive\footnote{https://www.cadc-ccda.hia-iha.nrc-cnrc.gc.ca/en/cfht/}
and both ESPaDOnS and Narval data can be retrieved from the 
PolarBase\footnote{http://polarbase.irap.omp.eu/} archive 
\citep{PolarBase}.

Additionally, one Potsdam Echelle Polarimetric and Spectroscopic Instrument
\citep[PEPSI;][]{pepsi2015,pepsi2018}
spectrum has been recorded in linear polarized light in 2017 October at the 
2$\times$8.4 m Large Binocular Telescope (LBT) on Mt.~Graham, Arizona and one 
unpolarized spectrum was also obtained with the Astrophysical Research 
Consortium Echelle Spectrograph 
\citep[ARCES;][]{arces}
on the ARC 3.5\,m telescope at the Apache Point Observatory (APO) on 2019 
March~1. 

\begin{table}
\centering
\caption{
 Logbook of the observations. The columns give the telescope and instrument
 configuration, the heliocentric Julian date (HJD), the exposure time, and 
 the $S/N$ around 6100\,\AA{}. HJDs followed by a $-$ or \textbar{} indicate 
 subsequent spectra that were combined to increase the $S/N$ in Stokes~$V$ 
 profiles that were used for the $\bz$ analysis. The HJD followed by a $\#$ 
 indicates the spectrum with the overall lowest $S/N$, and which was not used 
 in our measurements. HJDs followed by a $*$ indicate noisy spectra that were 
 not combined with other spectra obtained at similar epochs.
 }
\label{T:obs}
\begin{tabular}{llcc}
\noalign{\smallskip}\hline \noalign{\smallskip}
\multicolumn{1}{l}{Configuration} &
\multicolumn{1}{c}{HJD} &
\multicolumn{1}{c}{Exp.\ time} &
\multicolumn{1}{c}{$S/N$} \\
&
2\,450\,000+ &
[s] &
\\
\noalign{\smallskip}\hline \noalign{\smallskip}
CFHT + ESPaDOnS    & 55958.719 & 2400 & 109 \\
CFHT + ESPaDOnS    & 55959.719 & 2400 & 88 \\
CFHT + ESPaDOnS    & 55960.717 & 2400 & 71 \\
CFHT + ESPaDOnS    & 55961.717 & 2400 & 112 \\
CFHT + ESPaDOnS    & 55966.725 & 2400 & 90 \\
TBL  +  Narval     & 56011.330 & 4800 & 78 \\
CFHT + ESPaDOnS    & 56197.950$-$ & 5400 & 146 \\
CFHT + ESPaDOnS    & 56198.014$-$ & 5400 & 146 \\
CFHT + ESPaDOnS    & 56200.974 \textbar & 5400 & 130 \\
CFHT + ESPaDOnS    & 56201.039 \textbar & 5400 & 148 \\
CFHT + ESPaDOnS    & 56270.751$-$ & 5400 & 115 \\
CFHT + ESPaDOnS    & 56270.816$-$ & 5400 & 138 \\
CFHT + ESPaDOnS    & 56285.761$*$ & 5400 & 93 \\
CFHT + ESPaDOnS    & 56285.826 \textbar & 5400 & 127 \\
CFHT + ESPaDOnS    & 56285.890 \textbar & 5400 & 131 \\
CFHT + ESPaDOnS    & 56293.733$-$ & 5400 & 136 \\
CFHT + ESPaDOnS    & 56293.798$-$ & 5400 & 138 \\
CFHT + ESPaDOnS    & 56295.861$\#$ & 5400 & 19 \\
CFHT + ESPaDOnS    & 56532.126 & 5160 & 156 \\
CFHT + ESPaDOnS    & 56534.063 \textbar & 5160 & 154 \\
CFHT + ESPaDOnS    & 56534.124 \textbar & 5160 & 135 \\
CFHT + ESPaDOnS    & 56549.039$-$ & 5160 & 146 \\
CFHT + ESPaDOnS    & 56549.102$-$ & 5160 & 141 \\
CFHT + ESPaDOnS    & 56561.028 \textbar & 5160 & 150 \\
CFHT + ESPaDOnS    & 56561.091 \textbar & 5160 & 123 \\
CFHT + ESPaDOnS    & 56613.913$-$ & 5160 & 112 \\
CFHT + ESPaDOnS    & 56614.045$-$ & 5160 & 125 \\
CFHT + ESPaDOnS    & 56621.039 \textbar & 5160 & 140 \\
CFHT + ESPaDOnS    & 56621.102 \textbar & 5160 & 135 \\
CFHT + ESPaDOnS    & 56665.828$-$ & 5160 & 123 \\
CFHT + ESPaDOnS    & 56665.889$-$ & 5160 & 149 \\
CFHT + ESPaDOnS    & 57289.978$*$ & 5160 & 82 \\
CFHT + ESPaDOnS    & 57290.041 \textbar & 5160 & 106 \\
CFHT + ESPaDOnS    & 57291.059 \textbar & 5160 & 108 \\
LBT  + PEPSI       & 58042.022    & 3600 & 70 \\
ARC  + ARCES       & 58543.653    & 3600 & 98 \\
\noalign{\smallskip}\hline \noalign{\smallskip}
\end{tabular}
\end{table}

The logbook of all available observations is presented in Table~\ref{T:obs}.
The recorded linear polarization PEPSI Stokes~$Q$ and $U$ spectra and the ARCES 
spectrum were however very noisy, hence only the Stokes~$I$ spectrum was used 
for measuring the magnetic field modulus. The resolving power of all archival 
observations is $\sim 65\,000$, while the spectral resolution of the PEPSI 
spectrum is 130\,000 whereas the ARCES spectrum has a resolution of only 
31\,500. As most of the spectra have moderate signal-to-noise ratios ($S/N$) 
in the Stokes~$I$ spectra, mostly in the range of 100--150, some spectra 
obtained during the same night or, in one case, during subsequent nights have 
been combined to increase the signal for the Stokes~$V$ profiles. However, in 
cases where the $S/N$ of the spectrum was significantly lower than that of 
the other spectra obtained during the same night, only the spectra with 
better $S/N$ were used for the analysis of the mean longitudinal magnetic 
field. Some spectra with low $S/N$ have been included in the analysis of the 
line intensity variability or used for the measurements of the mean magnetic 
field modulus, if the Zeeman splitting was detectable on them. Further, to 
increase the accuracy of the mean longitudinal magnetic field determination, 
we applied the least-squares deconvolution technique (LSD) described in 
detail by 
\citet{Donati1997}.
To create line masks for the measurements, we have used the Vienna Atomic 
Line Database 
\citep[VALD; e.g.,][]{Kupka2011,VALD3}, 
adopting the stellar parameters of \mbox{NGC\,1624-2}, $\teff=35\,000$\,K 
and $\logg=4.0$ 
\citep{Wade2012a}. 
The mask for the photospheric absorption lines includes spectral lines 
belonging to the elements He, O, and C and is identical to that presented in 
Table\,7 in the work of 
\citet{Wade2012a}, 
apart from \ion{He}{i} $\lambda$7281, which is heavily distorted by numerous 
strong telluric lines in most of the spectra. In massive magnetic early 
B-type stars, He and Si are usually inhomogeneously distributed on the 
stellar surface, with the He abundance spots located close to the magnetic 
poles and the Si abundance spots close to the magnetic equator. Thus, to test 
a possible inhomogeneous distribution of Si, we also included in our line 
mask the \ion{Si}{iv} lines at $\lambda$4631 and $\lambda$4654. The available 
data set allows us, in addition, for the first time to study the variability 
of the high-excitation emission \ion{O}{III} lines and of the emission wind 
\ion{He}{i} lines over the rotation cycle. 

\section{Magnetic field measurements}
\label{sect:B}

\subsection{Mean longitudinal magnetic field $\bz$}

\begin{figure}
 \centering 
        \includegraphics[width=0.80\columnwidth]{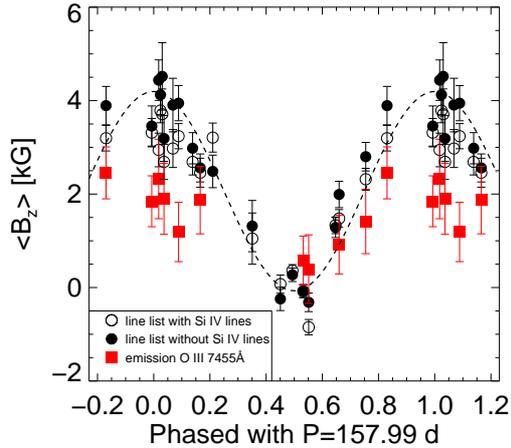}
        \caption{
         Mean longitudinal magnetic field $\bz$ measured using two different 
         line masks and using only the high-excitation line \ion{O}{iii} 
         $\lambda$7455 against the rotation phase. The dashed line represents 
         the best sinusoidal least-squares fit solution for the phase curve 
         defined by our magnetic field measurements (see text).
                }
   \label{fig:Bphase}
\end{figure}

\begin{table*}
\centering
\caption{
  Longitudinal magnetic field measurements of \mbox{NGC\,1624-2}. The dates 
  are as in Table~\ref{T:obs}, but with the average of the combined dates if 
  observations were co-added. The rotation phases are listed in Column~2, 
  followed by the measurements using the whole sample of photospheric 
  absorption lines and the sample without the \ion{Si}{iv} lines. In 
  Columns~5, 6, and 7 we list the measurements using the high-excitation 
  emission \ion{O}{iii} $\lambda$7455 line and the wind \ion{He}{i} 
  $\lambda$5876 and \ion{He}{i} $\lambda$6678 lines.}
\label{T:Bz}
\begin{tabular}{l c r@{$\pm$}l r@{$\pm$}l  r@{$\pm$}l r@{$\pm$}l r@{$\pm$}l}
\hline
\multicolumn{1}{c}{Date}            &
\multicolumn{1}{c}{Phase}            &
\multicolumn{10}{c}{$\bz$ (G)} \\
\multicolumn{1}{c}{}            &
\multicolumn{1}{c}{}            &
\multicolumn{2}{c}{ALL} &
\multicolumn{2}{c}{ALL $-$ Si} &
\multicolumn{2}{c}{\ion{O}{iii} $\lambda$7455}    &
\multicolumn{2}{c}{\ion{He}{i} $\lambda$5876} &
\multicolumn{2}{c}{\ion{He}{i} $\lambda$6678} \\
\hline
55958.719       & 0.018 & 2944    & 360    &  4438    & 433    &  2324    & 851        &  $-$2713   & 314  & $-$7420 & 791 \\
55959.719       & 0.024 & 3787    & 261    &  4125    & 425    &  \multicolumn{2}{c}{} &  $-$2538   & 294  & $-$6149 & 903 \\
55960.717       & 0.030 & 3710    & 540    &  4518    & 721    &  \multicolumn{2}{c}{} &  $-$1418   & 453  & $-$7050 & 1050 \\
55961.717       & 0.037 & 2689    & 376    &  3189    & 490    &  1899    & 761        &  $-$1818   & 259  & $-$6838 & 800 \\
55966.724       & 0.068 & 2975    & 406    &  3905    & 573    &  \multicolumn{2}{c}{} &  $-$2869   & 401  & $-$4837 & 1066 \\
56011.330       & 0.351 & 1048    & 548    &  1315    & 549    &  \multicolumn{2}{c}{} &  $-$820    & 1016 & $-$2417 & 1813 \\
56197.982       & 0.532 & $-$80   & 137    &  $-$74   & 139    &  578     & 524        &  1674      & 224  & $-$65   & 468 \\
56201.006       & 0.551 & $-$848  & 164    &  $-$316  & 199    &  394     & 735        &  182       & 336  & 312     & 508 \\
56270.783       & 0.993 & 3314    & 304    &  3456    & 430    &  1846    & 544        &  $-$2550   & 235  & $-$4074 & 750 \\
56285.858       & 0.088 & 3239    & 269    &  3945    & 376    &  1189    & 635        &  $-$3201   & 185  & $-$4074 & 523 \\
56293.765       & 0.138 & 2688    & 279    &  2981    & 334    &  \multicolumn{2}{c}{} &  $-$2176   & 263  & $-$4121 & 595 \\
56532.126       & 0.647 & 1332    & 178    &  1284    & 215    &  \multicolumn{2}{c}{} &  $-$1792   & 504  & $-$1097 & 546 \\
56534.093       & 0.659 & 1478    & 190    &  1992    & 282    &  920     & 629        &  $-$2150   & 379  & $-$ 309 & 569 \\
56549.070       & 0.754 & 2317    & 234    &  2800    & 306    &  1415    & 690        &  $-$3524   & 232  & $-$3669 & 540 \\
56561.059       & 0.830 & 3195    & 273    &  3893    & 411    &  2454    & 556        &  $-$3727   & 265  & $-$2431 & 567 \\
56613.979       & 0.165 & 2457    & 310    &  2559    & 300    &  1872    & 723        &  $-$5120   & 488  & $-$5240 & 783 \\
56621.070       & 0.210 & 3209    & 311    &  2485    & 348    &  \multicolumn{2}{c}{} &  $-$7846   & 460  & $-$3324 & 628 \\
56665.858       & 0.493 & 363     & 128    &  272     & 144    &  \multicolumn{2}{c}{} &  $-$480    & 897  & 15      & 697 \\
57290.550       & 0.447 & 73      & 195    &  $-$244  & 249    &  \multicolumn{2}{c}{} &  155       & 1058 & $-$620  & 732 \\
\hline
\end{tabular}
\end{table*}

The mean longitudinal magnetic field is determined by computing the 
first-order moment of the LSD Stokes~$V$ profile according to
\citet[][]{Mathys1989}:

\begin{equation}
\left<B_{\mathrm z}\right> = -2.14 \times 10^{11}\frac{\int \upsilon V
  (\upsilon){\mathrm d}\upsilon }{\lambda_{0}g_{0}c\int
  [I_{c}-I(\upsilon )]{\mathrm d}\upsilon},
\end{equation}

\noindent
where $\upsilon$ is the velocity shift from the line centre, in \kms, and 
$\lambda_{0}$ and $g_{0}$ are the normalization values of the line weights. 
We have adapted for them here the average values of, respectively, the 
wavelengths (in nm) and the effective Land\'e factors of all the lines used 
to compute the LSD profile. The results of our magnetic field measurements 
using the line mask for the photospheric absorption lines, including the Si 
lines, the line mask without the Si lines, and the measurements using the 
high excitation emission line \ion{O}{iii} $\lambda$7455 and the strongest 
emission lines \ion{He}{i} $\lambda$5876 and \ion{He}{i} $\lambda$6678 are 
presented in Fig.~\ref{fig:Bphase} and in Table~\ref{T:Bz} in Columns~3 to 7.
All LSD profiles for the selected line masks, as well as Stokes~$I$, $V$, 
and null profiles for the individual emission lines are shown in 
Figs.~\ref{afig:IVNabsl}--\ref{afig:IVN7065} in the Appendix. The average 
effective Land\'e factors used for the normalization for the line mask 
including the Si lines and the mask without the Si lines are 1.55 and 1.24, 
respectively. The corresponding wavelengths are 5013 and 5224\,\AA. For the 
profiles calculated using all lines and the line list without the Si lines an 
integration was made from $-120$ to $+25$\,\kms, for \ion{O}{iii} 
$\lambda$7455 from $-70$ to $+10$\,\kms, for \ion{He}{i} $\lambda$5876 and 
for \ion{He}{i} $\lambda$6678 from $-200$ to $+150$\,\kms.

In most cases the false alarm probability (FAP) for the field detections is 
less than $10^{-6}$. According to 
\citet{Donati1992}, 
a Zeeman profile with FAP $\leq 10^{-5}$ is considered as a definite 
detection, $10^{-5} <$ FAP $\leq 10^{-3}$ as a marginal detection, and FAP 
$> 10^{-3}$ as a non-detection. Our measurements show that the $\bz$ values 
become lower if the \ion{Si}{iv} lines are included in the line mask. A 
possible scenario to explain these measurements would be an inhomogeneous 
surface element distribution with a region of Si concentration located closer 
to the magnetic equator. It is also possible that we observe a filling of the 
lines by emission. Our $\bz$ measurements in the range from $-0.2$ to 4.5\,kG 
obtained for the sample of the photospheric absorption lines are in good 
agreement with the $\bz$ values reported by 
\citet{Grunhut2017}, 
but disagree with the measurement of $\bz=5.35$\,kG reported by 
\citet{Wade2012a}. 
Assuming a limb-darkening coefficient $u=0.3$
\citep{Claret2011} 
and a centred dipolar magnetic field, we estimate the dipole strength of 
\mbox{NGC\,1624-2} $B_{\rm d}\ge16$\,kG according to Equation~(1) of 
\citet{Preston1967}. 
The dashed line in Fig.~\ref{fig:Bphase} represents the best sinusoidal 
least-squares fit solution for the phase curve defined by our magnetic field 
measurements using the line mask for photospheric absorption lines, without 
the \ion{Si}{iv} lines. The rotation phase presented in Table~\ref{T:Bz} in 
the second column is calculated using the rotation period of 157.99\,d 
adopted by 
\citet{Wade2012a} 
and the initial epoch for the phase origin HJD$_{0} = 2455955.94\pm0.21$ 
corresponding to the maximum of the longitudinal field phase curve. The 
fitted sinusoid has an amplitude $A_{\left<B_{\rm z}\right>} = 2132\pm364$\,G 
and a mean value $\overline{\left<B_{\rm z}\right>} = 2063\pm328$\,G, thus
$\bz_{\rm max}=4195$\,G and $\bz_{\rm min}=-69$\,G.

\begin{figure*}
 \centering 
 \includegraphics[width=0.245\textwidth]{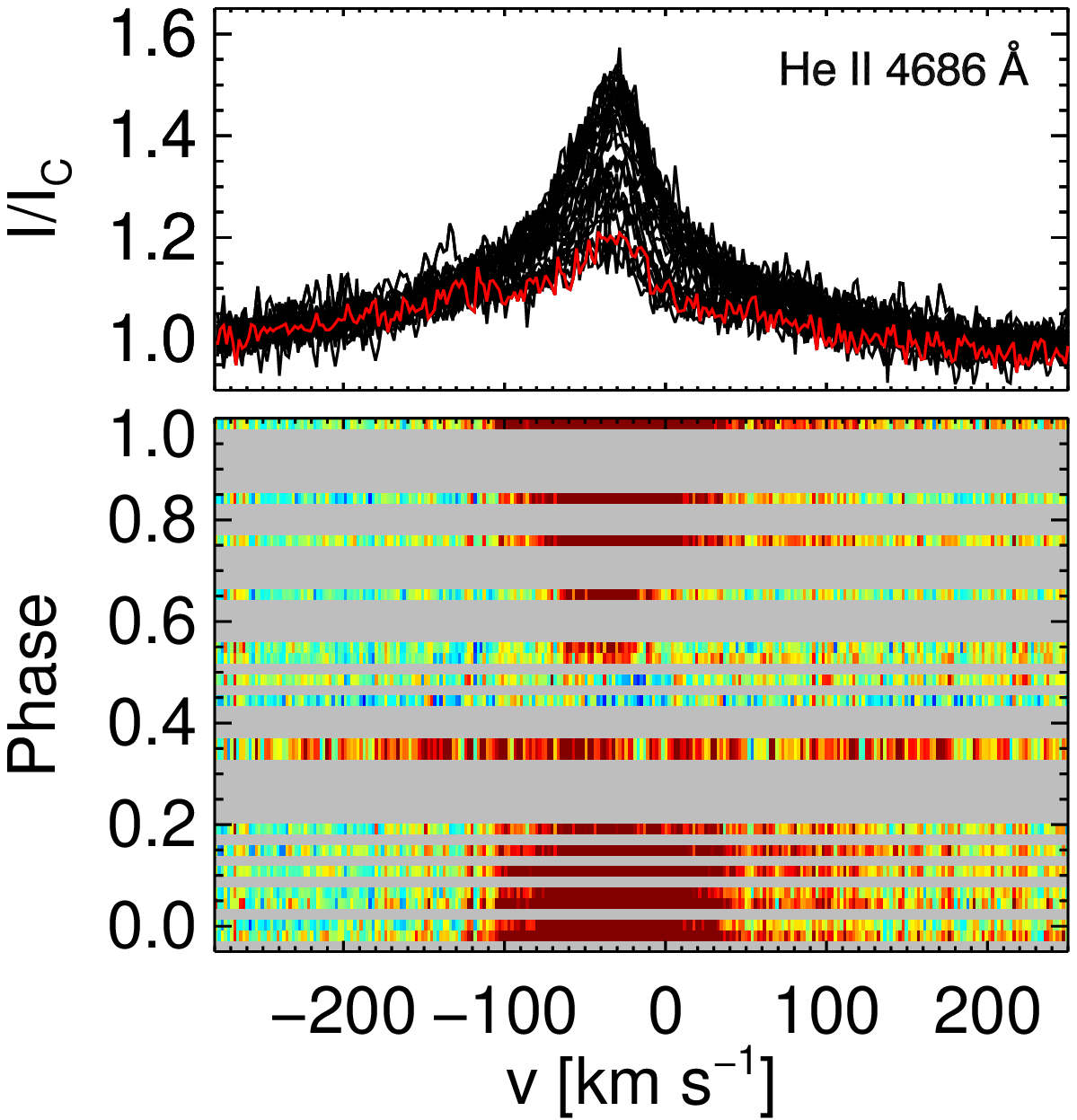}
 \includegraphics[width=0.245\textwidth]{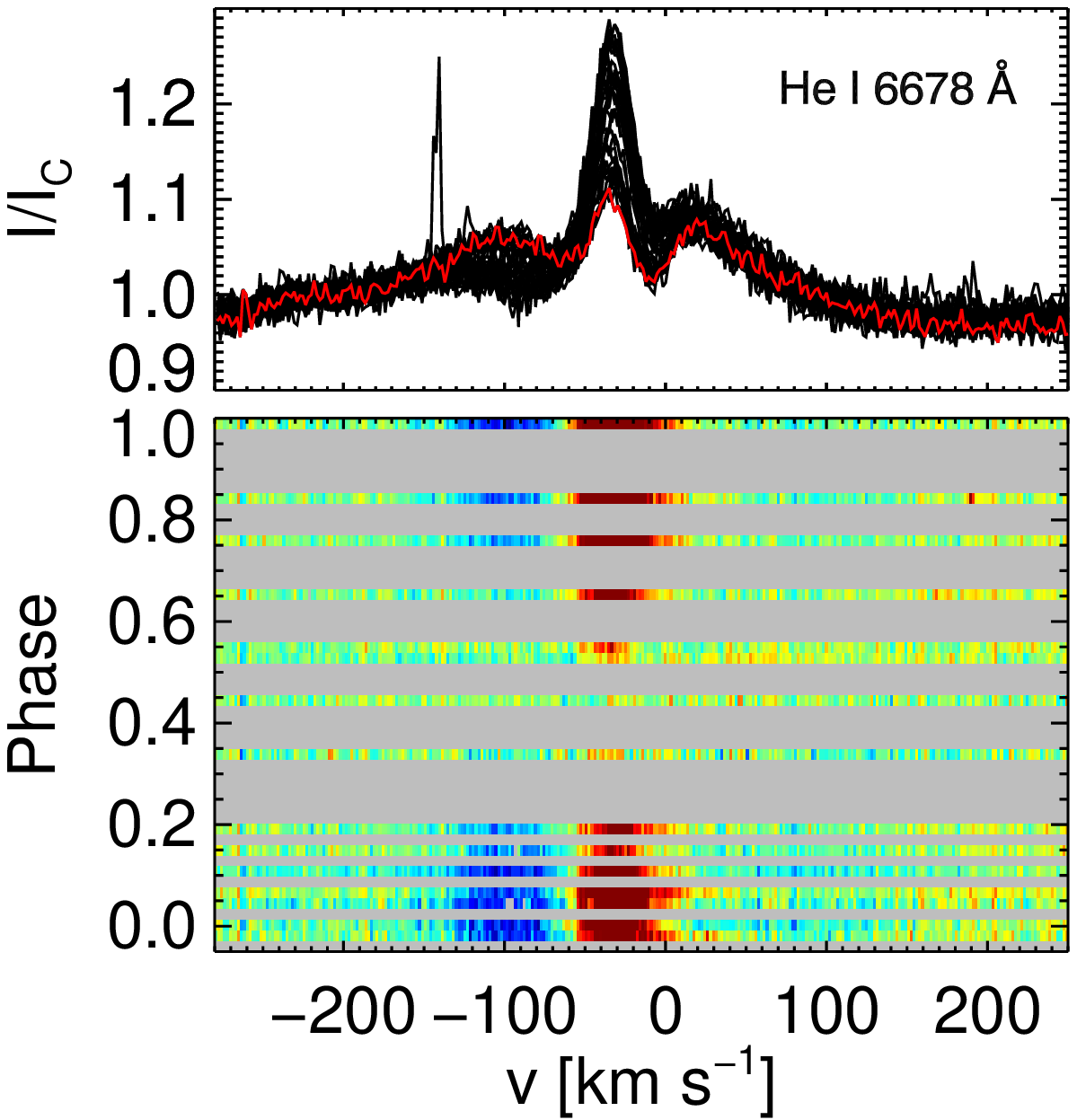}
 \includegraphics[width=0.245\textwidth]{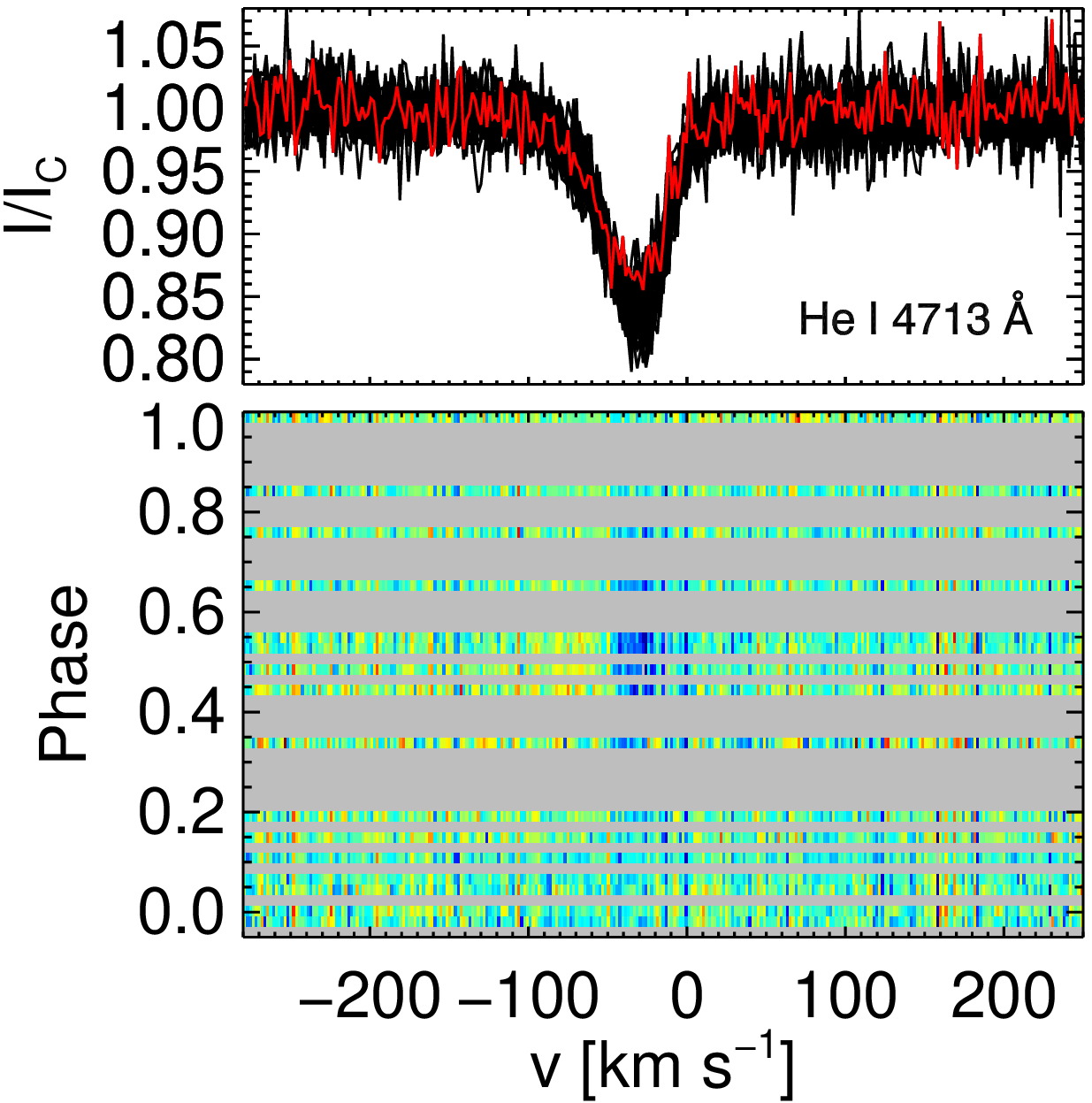}
 \includegraphics[width=0.245\textwidth]{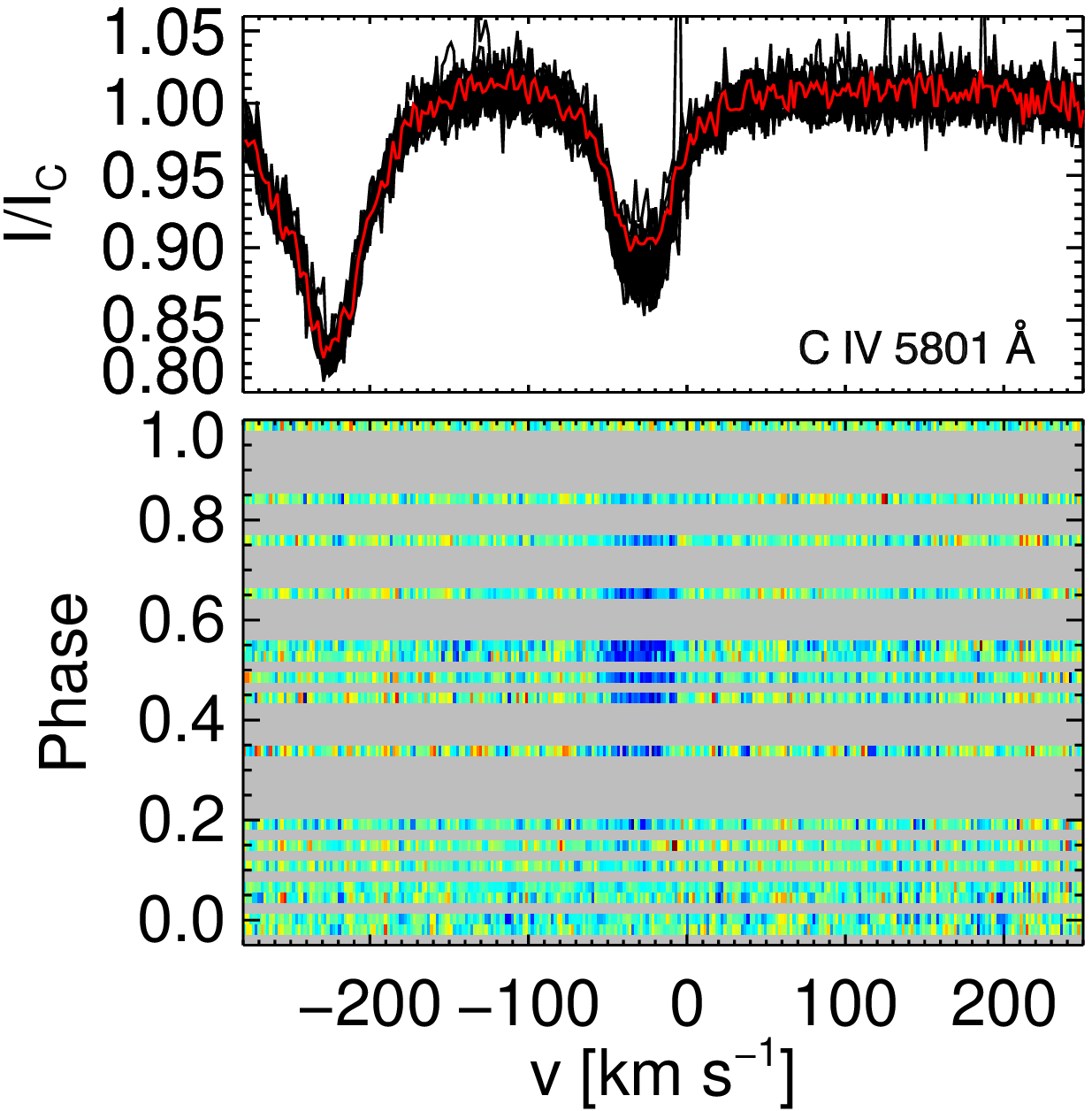}
        \caption{Examples of line intensity variability of the emission lines 
	  \ion{He}{ii} $\lambda$4686 and 
          \ion{He}{i} $\lambda$6678, and the photospheric absorption lines
          \ion{He}{i} $\lambda$4713 and
          \ion{C}{iv} $\lambda$5801 as a function of the rotation phase.
          In the upper panel we present all profiles overplotted with the 
          observed weakest profiles highlighted in red colour.
          The lower panels show the dynamic spectra of the difference 
          between the individual line profiles and the weakest emission or 
          absorption profile.
                  }
   \label{fig:dyn4686}
\end{figure*}

Knowing the rotation period, we can estimate the equatorial velocity and the 
inclination of the rotation axis of the star $i$. Assuming a radius 
$R=9.7\,{\rm R}_{\odot}$ 
\citep{Petit2013}
and a projected rotation velocity $v\,\sin\,i\le3$\,\kms{} 
\citep{Wade2012a},
we obtain $i\le75^\circ$. Using the well known relations developed by 
\citet{Stibbs1950}
and 
\citet{Preston1967}
for a centred magnetic dipole tilted to the rotation axis by angle $\beta$,
we estimate an obliquity angle $\beta \ge 15.5^{\circ}$. However, with 
$v\,\sin\,i = 1$\,km\,s$^{-1}$, we obtain $i = 19^{\circ}$ and 
$\beta = 72^{\circ}$, i.e.\ nearly a flip of the two axes, showing that the 
exact geometry of the magnetic field can not be determined with the still too 
loose constraint on $v\,\sin\,i$. The measured values of the longitudinal 
magnetic field of \mbox{NGC\,1624-2} with predominantly positive polarity 
imply that only one magnetic pole is observable over the rotation cycle. The 
dynamic spectra of emission and photospheric lines show a single-wave 
variation of their intensity: the intensity of the lines H$\alpha$, 
\ion{He}{ii} $\lambda$6486, \ion{He}{i} $\lambda$5876 and $\lambda$6678 are 
the strongest close to the positive extremum of $\bz$ whereas the 
photospheric absorption lines of \ion{He}{i}, \ion{C}{iv}, \ion{Si}{iv}, and 
the emission \ion{O}{iii} lines are the strongest close to the rotation phase 
0.5. Examples of the variability of different lines are presented in 
Fig.~\ref{fig:dyn4686}.

\subsection{Mean magnetic field moduls $\langle B \rangle$}

\begin{table}
\centering
\caption{
 The mean magnetic field modulus $\langle B \rangle$ values obtained in the 
 spectra with measurable splitting in the \ion{C}{iv} $\lambda$5812 line.
 }
\label{T:Bmod}
\begin{tabular}{l c r@{$\pm$}l}
\hline
\multicolumn{1}{c}{Date}            &
\multicolumn{1}{c}{Phase}            &
\multicolumn{2}{c}{$\langle B \rangle$ (G)} \\
\noalign{\smallskip}\hline \noalign{\smallskip}
 55959.719 & 0.024 &   9731   & 431     \\
 55960.717 & 0.030 &   9433   & 526     \\
 55961.717 & 0.037 &   9650   & 950     \\
 55966.725 & 0.068 &   10927  & 300     \\
 56011.330 & 0.351 &   11546  & 1053     \\
 56197.950 & 0.532 &   11424  & 304     \\
 56198.014 & 0.532 &   10975  & 409     \\
 56200.974 & 0.551 &   11059  & 356     \\
 56201.039 & 0.551 &   10990  & 360     \\
 56270.751 & 0.993 &   10299  & 385     \\
 56270.816 & 0.993 &   10973  & 491     \\
 56285.826 & 0.088 &   9253   & 713     \\
 56285.890 & 0.088 &   9682   & 423     \\
 56293.733 & 0.138 &   10322  & 616     \\
 56293.798 & 0.138 &   9932   & 799     \\
 56532.126 & 0.647 &   11099  & 301     \\
 56534.063 & 0.659 &   11384  & 310     \\
 56534.124 & 0.660 &   11405  & 377     \\
 56549.039 & 0.754 &   11586  & 497     \\
 56549.102 & 0.754 &   11372  & 408     \\
 56561.028 & 0.830 &   10616  & 379     \\
 56613.913 & 0.165 &   10751  & 485     \\
 56614.045 & 0.165 &   9127   & 950     \\
 56621.039 & 0.210 &   10433  & 713     \\
 56621.102 & 0.210 &   9510   & 475     \\
 56665.828 & 0.493 &   10840  & 341     \\
 56665.889 & 0.494 &   10803  & 346     \\
 57289.978 & 0.444 &   11396  & 319     \\
 57290.041 & 0.444 &   10886  & 294     \\
 57291.059 & 0.451 &   11683  & 555     \\
 58042.022 & 0.204 &   11122  & 600     \\
 58543.653 & 0.379 &   10913  & 425     \\
\noalign{\smallskip}\hline \noalign{\smallskip}
\end{tabular}
\end{table}


\begin{figure}
 \centering 
        \includegraphics[width=0.80\columnwidth]{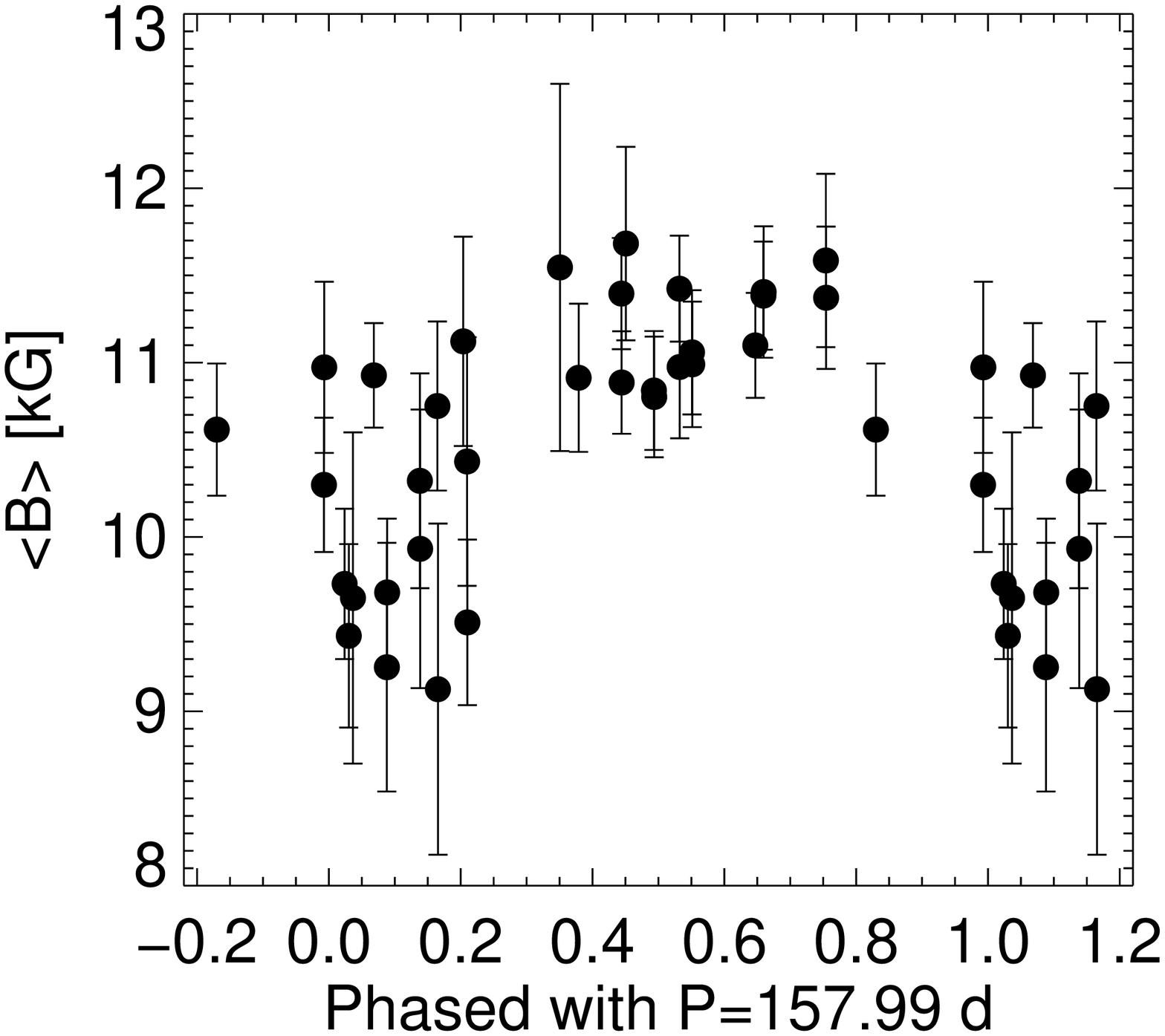}
        \caption{
	  Mean magnetic field modulus $\langle B \rangle$ of 
          \mbox{NGC\,1624-2} measured using the split components of the 
          magnetically resolved line \ion{C}{iv} $\lambda$5812 against the 
          rotation phase.
                  }
   \label{fig:Bmod}
\end{figure}

The availability of the variation curve of the mean magnetic field modulus 
$\langle B \rangle$ allows us to put stronger constraints on the structure of 
the magnetic field of \mbox{NGC\,1624-2} as compared to any other magnetic 
O-type star with unsplit lines. Our measurements of $\langle B \rangle$, 
which is the average over the visible stellar hemisphere of the modulus of 
the magnetic field vector, weighted by the local line intensity, are 
presented in Table~\ref{T:Bmod} and Fig.~\ref{fig:Bmod}. The field modulus 
was calculated by fitting the magnetically resolved \ion{C}{iv} $\lambda$5812 
line with two Gaussian profiles and using the relation given e.g.\ by 
\citet{HubrigNesvacil2007}.
The uncertainties in the presented $\langle B \rangle$ values are due to the 
accuracy of the measured wavelengths of the blue and red split line components.
Our measurements indicating a field intensity between 9 and 12\,kG differ from 
the measurements reported by 
\citet{Wade2012a}, 
who reported $\langle B \rangle = 15$\,kG for the $\lambda$5801 line and  
$\langle B \rangle = 13$\,kG for the $\lambda$5812 line using the Narval 
spectrum. This difference can likely be explained by the Zeeman pattern 
of the \ion{C}{iv} $\lambda$5812 line. We interpret in our measurements the 
splitting as a pseudo-doublet, which underestimates the strength of the field 
modulus.

The approximately sinusoidal variation of $\bz$ and the ratio of the values 
of the extrema of $\langle B \rangle$ indicates that there is an important 
component of the field that is dipolar. Further, a comparison of the 
variation curves of the mean magnetic field modulus and the mean longitudinal 
magnetic field shows that the minimum of the mean magnetic field modulus 
corresponds to the maximum of the mean longitudinal magnetic field. This 
indicates that the field structure must significantly depart from a centred 
dipole. It is possible that we see a small shift between the modulus minimum 
with respect to the positive extremum of the $\bz$, although the fidelity of 
such a shift is only barely supported, given the rather large dispersion of 
the data points. A future confirmation of the presence of a shift is 
necessary, as this would indicate that the magnetic field structure of 
\mbox{NGC\,1624-2} slightly departs from a simple dipole. 

\subsection{Emission lines}

We confirm the conclusion of 
\citet{Wade2012a} 
that the $\bz$ values measured using the high-excitation \ion{O}{iii} 
emission lines, suggested to form in the low-velocity plasma confined in 
closed magnetic loops above the stellar surface, are significantly lower than 
those obtained for photospheric absorption lines. The $\bz$ values obtained 
using the \ion{O}{iii} $\lambda$7455 line are in the range from 0.4 to 
2.3\,kG and show a variability pattern similar to that detected for the 
absorption lines.

Interestingly, we detect in the spectra the forbidden [\ion{O}{i}] 6300 and 
6363 lines, indicating the presence of low-density atomic material in the 
vicinity of \mbox{NGC\,1624-2}. The formation mechanism of these lines is 
unclear. It was suggested that they originate from disk winds or regions 
where stellar UV radiation impinges on the disk surface in pre-main 
sequence Herbig Ae/Be stars 
\citep[e.g.][]{Finkenzeller, Corcoran}. 
It is also possible that the forbidden [\ion{O}{i}] 6300 and 6363 lines are 
formed in magnetospheric accretion columns 
\citep{Muzerolle}.
The variability of the strength of these lines is presented in 
Fig.~\ref{bfig:nebular}. They are the strongest at the rotational phases 
0.03 and 0.35 and do not show rotational modulation. This suggests that they 
are most likely formed beyond the Alfv\'en radius.

Our analysis of the observations of \mbox{NGC\,1624-2}, in the scenario of an 
oblique magnetic rotator, indicates that stronger emissions in the hydrogen and 
helium emission lines are detected when the magnetically-confined cooling 
disk is seen closer to face-on, while the emission contribution is reduced 
when the cooling disk is seen closer to edge-on (see Fig.~\ref{fig:dyn4686}).

\begin{figure}
 \centering 
        \includegraphics[width=0.235\textwidth]{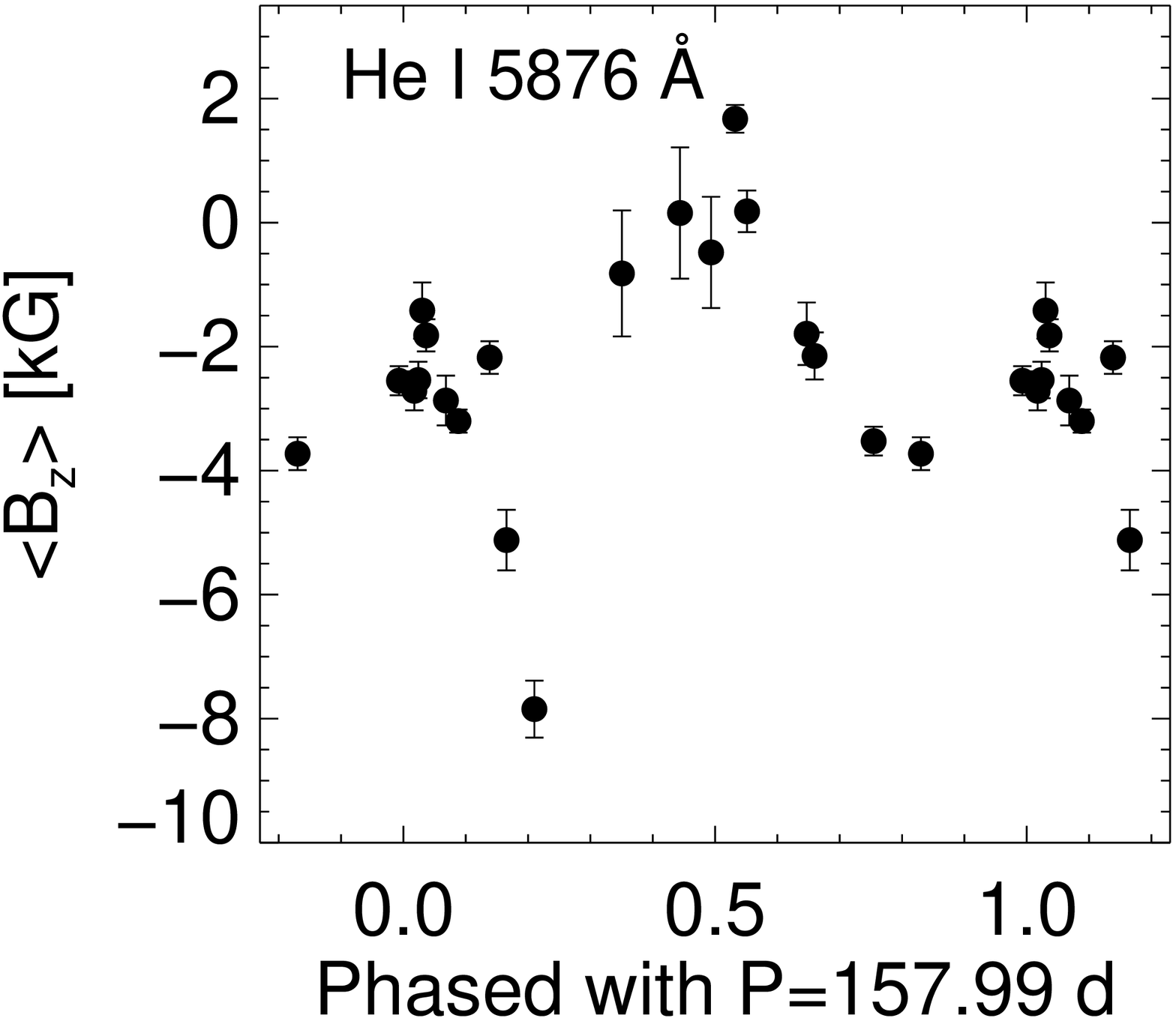}
        \includegraphics[width=0.235\textwidth]{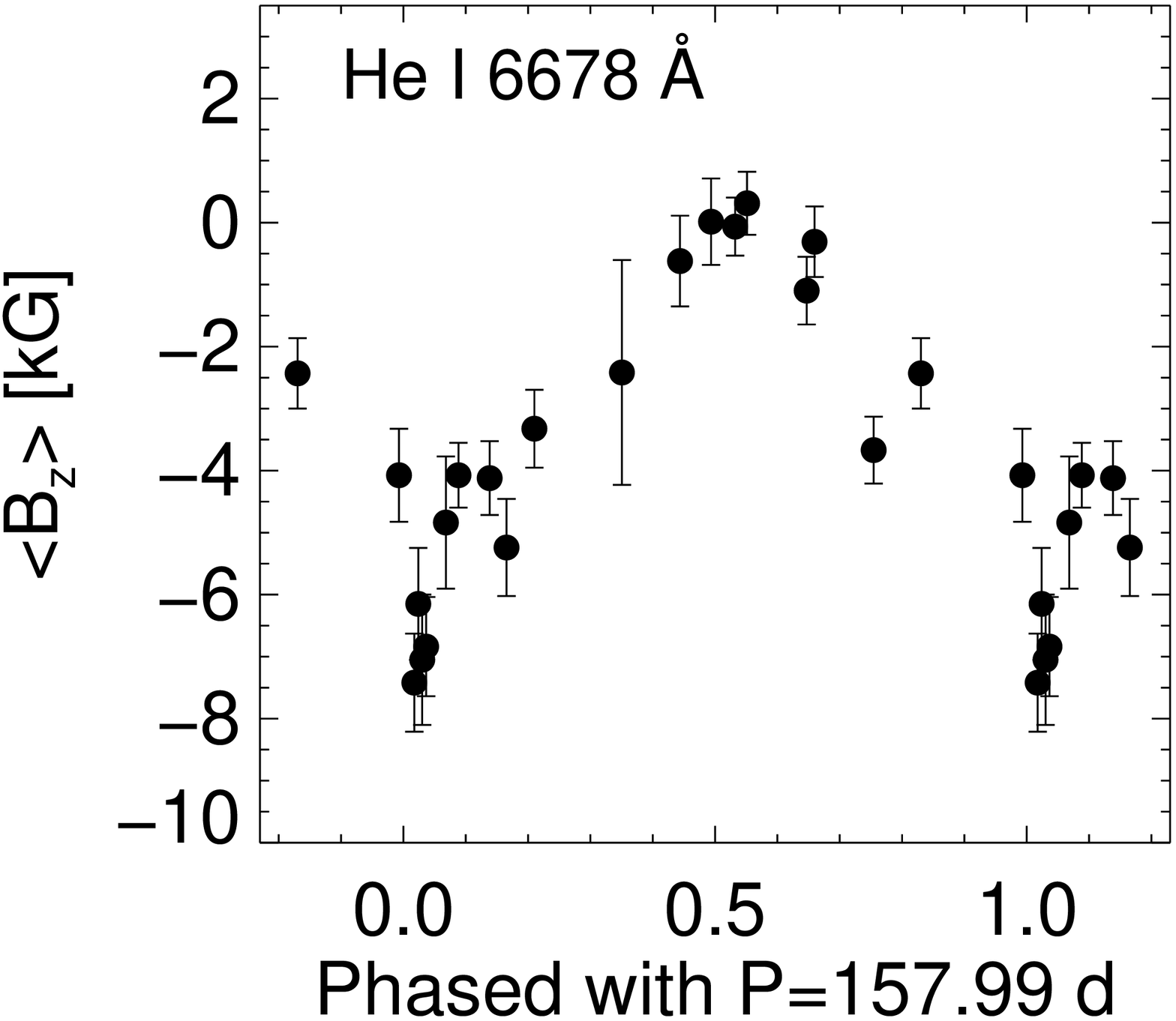}
        \caption{
         As Fig.~\ref{fig:Bphase}, but $\bz$ measured using the 
         \ion{He}{i} 5876\,\AA{} and \ion{He}{i} 6678\,\AA{} lines.
        }
   \label{fig:HeBphase}
\end{figure}

No measurements using the \ion{He}{i} emission lines were presented in 
the study of 
\citet{Wade2012a}. 
Our $\bz$ values are predominantly negative and, similar to the measurements 
of other spectral lines, show clear variability over the rotation period, 
with the negative $\bz$ extremum observed around phase 0, where the 
photospheric absorption lines show a positive $\bz$ extremum. The 
distribution of our measurement values over the rotation period obtained 
using the \ion{He}{i} $\lambda$5876 and $\lambda$6678 lines, which are the 
strongest in the $2^3P^0-n^3D$ and the $2^1P^0-n^1D$ series, respectively,
is presented in Fig.~\ref{fig:HeBphase}. The significant data point 
dispersion and the presence of two outliers in the measurements of 
\ion{He}{i} $\lambda$5876 close to the rotation phase 0.2 is caused by the 
short-term variability of the line intensity on a time scale of several days,
and is likely a consequence of either the pulsational variability recently 
detected by 
\citet{Kurtz} 
or a time-variable structure of gas flows within the magnetosphere
\citep{udDoula2013}.

The origin of the emission in the \ion{He}{i} lines in the red spectral 
region is not well understood. A long time ago, 
\citet{Auer} 
showed that NLTE effects are quite small for \ion{He}{i} lines in the 
blue-violet region of the spectrum, whereas the red \ion{He}{i} lines, in 
particular \ion{He}{i} $\lambda$5876 and \ion{He}{i} $\lambda$6678, are 
strongly affected by departures from LTE. A large discrepancy between 
observations and their NLTE results occurred for \ion{He}{i} $\lambda$5876, 
where the observed lines are consistently stronger than computed. However, 
the presented NLTE calculations always remained in absorption. Only at the 
highest temperatures of about 50\,000\,K the NLTE results predicted that 
lines come into emission. 

Several mechanisms were suggested since then to understand the appearance of 
the emission observed in these lines. To explain the emissions in the 
\ion{He}{i} $\lambda$5876, $\lambda$6678 lines in the early type star 
$\lambda$~Eri,
\citet{Smithetal}
suggested that emission can be produced in a dense slab that is optically 
thick in the Lyman continuum. To produce a large enough column density for an 
observable feature, the requirement was made that the slab must be cohesive
and is confined in a magnetic loop to resist the acceleration of the ambient 
wind.

\begin{figure}
 \centering 
        \includegraphics[width=0.95\columnwidth]{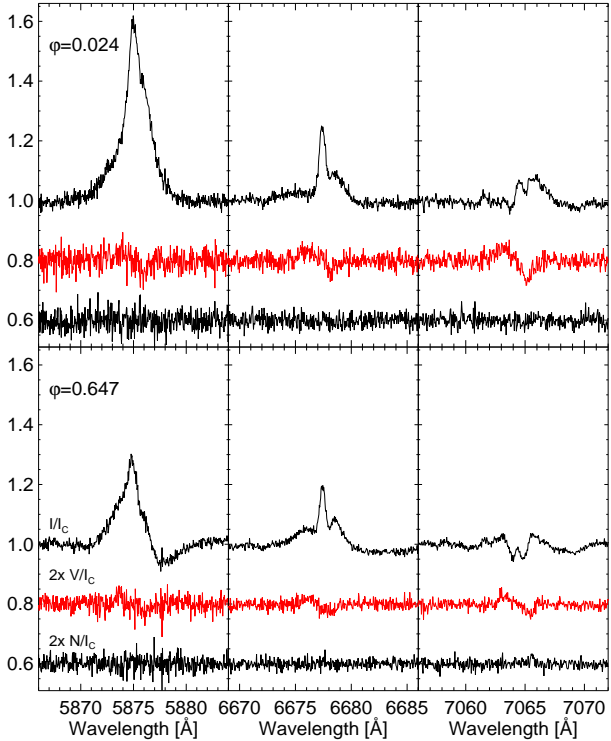}
        \caption{
	Examples of Stokes~$I$ (top), Stokes~$V$ (middle, shifted to 0.8
        and multiplied by two) and null (bottom, shifted to 0.6) profiles of 
        the emission lines 
	\ion{He}{i} $\lambda$5876, \ion{He}{i} $\lambda$6678, and 
	\ion{He}{i} $\lambda$7065 at two phases.
}
   \label{fig:windIVN}
\end{figure}

\begin{figure}
 \centering 
        \includegraphics[width=0.95\columnwidth]{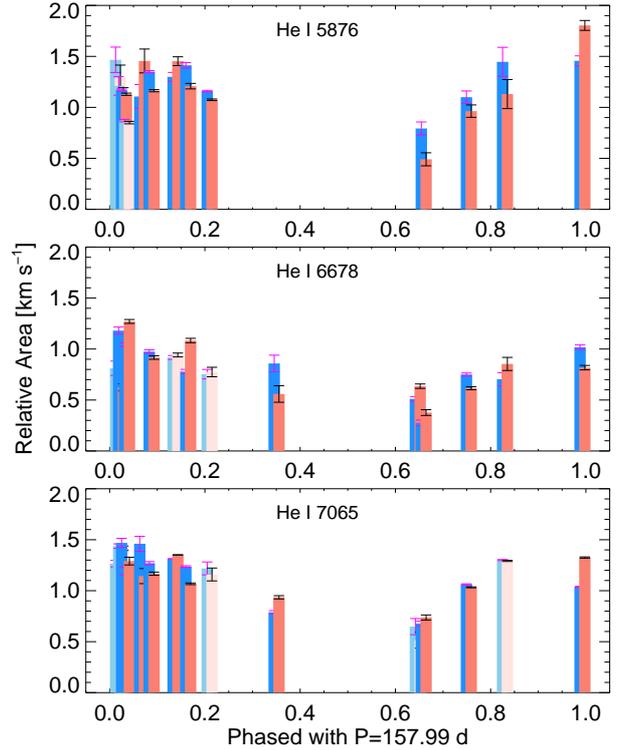}
        \caption{
        Differences between the relative areas of the blue and red parts of 
        the Stokes~$V$ profiles calculated at different rotation phases for the 
        emission lines \ion{He}{i} $\lambda\lambda$5876, 6678, and 7065.
	The lighter color represents the cases where the differences in the 
        areas are within the measurement uncertainties and the darker color 
        the cases where they are larger than the measurement uncertainties. 
        Error bars for each calculated blue and red part of the Stokes~$V$ 
        profiles are presented at the top of the individual bars.
}
   \label{fig:Vhisto}
\end{figure}

The presence of variable Zeeman features corresponding to a longitudinal 
field of negative polarity observed in the \ion{He}{i} emission lines, while 
high-excitation emission \ion{O}{iii} lines show the same polarity as the
photospheric absorption lines, is puzzling and currently difficult to 
interpret. One alternative hypothesis could be that the \ion{He}{i} 
Stokes~$I$ and $V$ profiles have a composite structure and present a 
contribution of several components. In Fig.~\ref{fig:windIVN} we present the 
Stokes~$I$, $V$, and null profiles for all three \ion{He}{i} emission lines 
at two rotational phases. Indeed, depending on the line, the observed 
Stokes~$I$ profiles indicate a rather complex variable structure consisting 
of at least two or three components (see also 
Figs.~\ref{afig:IVN5875}--\ref{afig:IVN7065}). It is quite possible that 
there is a correspondence between the different parts of the Stokes~$I$ 
profiles and those of the Stokes~$V$ profiles. The resulting composite 
Stokes~$V$ profiles would then present just a blend of the individual 
Stokes~$V$ profiles corresponding to different parts of the Stokes~$I$ 
profiles. It is noteworthy that for the majority of the obtained 
observations, the shape of the Stokes~$V$ profiles displayed in 
Figs.~\ref{afig:IVN5875}--\ref{afig:IVN7065} cannot be explained in terms 
of the Zeeman effect: if these profiles are indeed formed in a hydrostatic 
stellar atmosphere lacking gradients in velocity and magnetic field as a 
function of optical depth
\citep[e.g.][]{sanchez, lopez}, 
they always have the zero-order moment equal to zero, that is the integral of 
Stokes~$V$ over the region of the spectral line in Stokes~$I$ is equal to 
zero. The observed differences between the areas corresponding to the blue 
and red parts of the Stokes~$V$ profiles calculated for all three \ion{He}{i} 
emission lines are presented in Fig.~\ref{fig:Vhisto}. The differences were 
calculated only in rotation phases with a lower impact of the noise on the 
shape of the Stokes~$V$ profiles. As this figure shows, for all three lines 
the differences between the areas corresponding to the blue and red parts of 
the Stokes~$V$ profiles are significant. 

The changes in emission strength of the He lines and the different sign of 
the longitudinal magnetic field in these lines can also be understood within 
the following scenario. For a dipole magnetic field, the magnetic field lines 
reverse their orientation at the equator relative to the pole. If emitting 
material accummulates in the magnetosphere at the magnetic equator, it 
experiences a magnetic field of opposite orientation to that at the magnetic 
pole. Thus the longitudinal magnetic field measured in the \ion{He}{i} lines 
as seen in emission from the magnetospheric material at the magnetic equator 
would have a sign opposite from the longitudinal field measured from a 
photospheric absorption line. The extended region near the equator with an 
elevated density seems like a plausible source of circumstellar emission (see 
Fig.~\ref{fig:sim1} for an illustration). The magnetic field there has an 
opposite orientation to that dominating the upper hemisphere of the 
photosphere. Then, as the star rotates, the longitudinal component of the 
magnetic field in the emitting region becomes weaker, and the projected area 
of the emitting region diminishes, and possibly becomes partially occulted 
by the star.

Another alternative hypothesis could be related to the complex structure of the 
magnetospheric circumstellar environment of \mbox{NGC\,1624-2} and the possible 
presence of a reconnection-driven emission process. To understand the 
behaviour of the gas flow and the mass distribution within the magnetosphere, 
we carried out first MHD numerical simulations for this star.

\section{MHD numerical simulations}

A dipolar magnetic field with a polar field strength of 16\,kG will strongly 
alter the gas flow from the star and inhibit mass loss. The impact of the 
magnetic field is described by the wind confinement parameter, 
$\eta_* = B_{\rm eq}^2R_*^2/\dot{M}\varv_\infty$, where $B_{\rm eq}$ is the 
field strength on the stellar surface at the magnetic equator, $R_*$ the 
stellar radius, $\dot{M}$ the mass loss rate with no magnetic field, and 
$\varv_\infty$ the terminal velocity of the stellar wind 
\citep{2002ApJ...576..413U}. 
For NGC 1624-2, $\eta_*=1.5 \times 10^4$ was found by
\citet{Wade2012a}, 
which results in a value of $11.3\,R_*$ for the Alfv\'en radius, which 
indicates the extent of the magnetically-dominated region. Within that 
region, the magnetic field will keep its dipole geometry and gas flowing from 
the star at low to mid-latitudes will be trapped. Only gas originating from 
the polar caps leaves the system unimpeded along open field lines.

To study the field geometry, mass distribution, and gas motions in the 
vicinity of the star, we ran a series of numerical simulations. The general 
model setup is described in 
\citet{2017AN....338..868K}. 
We used the {\sc NIRVANA} MHD code 
\citep{2008CoPhC.179..227Z} 
in spherical polar coordinates. The geometry is axisymmetric with respect to 
the polar axis of the magnetic field and the gas is assumed isothermal, with 
the gas temperature being the effective temperature of the star. No rotation 
is included. As the model does not include cooling, magnetic diffusion has 
been added with a diffusion coefficient of $10^{15}{\rm cm^2 s^{-1}}$.
The stellar and wind parameters have been adopted from 
\citet{Wade2012a}. 
We thus use a stellar mass of 34 solar masses, a stellar radius of 10 solar 
radii and an effective temperature of 35\,000\,K. The mass loss rate is
$1.6 \times 10^{-7} M_\odot  {\rm yr^{-1}}$  and the terminal wind velocity 
is 2785\,\kms{}. The start solution is a dipole magnetic field and a velocity 
law given by
\begin{equation}
   \varv(r) = \varv_\infty \left ( 1-\frac{R_*}{r}\right). \label{v0}
\end{equation} 
The initial mass distribution is determined by the velocity law (\ref{v0}) 
and the mass loss rate, i.e.\
\begin{equation}
  \rho=\frac{\dot{M}}{4 \pi \varv r^2}.
\end{equation}

A radiative force term after 
\citet{1975ApJ...195..157C} 
with $\bar{Q}=550$ and $\alpha=0.55$ is used to drive the wind. The parameters 
have been chosen such that without the magnetic field a stationary solution 
with the correct wind parameters is produced. We then repeat the computation 
with a dipole field included. To limit the cost in CPU time, the outer 
boundary of the simulation box is placed at 5 stellar radii and a polar field 
strength of 4\,kG is chosen. While this is smaller than the observed field 
strength of 16\,kG, it still leads to an extended region of magnetic 
confinement with $\eta_* \approx 700$ and an Alfv\'en radius of 5.4 stellar 
radii. Within the simulation box, the magnetic field geometry stays close to 
the original dipolar geometry and only a fraction of the gas flowing from the 
stellar surface can escape the system. The mass loss rate found by 
\citet{Wade2012a} 
is based on a prescription by 
\citet{2001A&A...369..574V} 
that does not take the magnetic field into account. It therefore refers to 
the amount of gas that would leave the system if there was no magnetic field.

\begin{figure}
\centering
    \includegraphics[width=.23\textwidth]{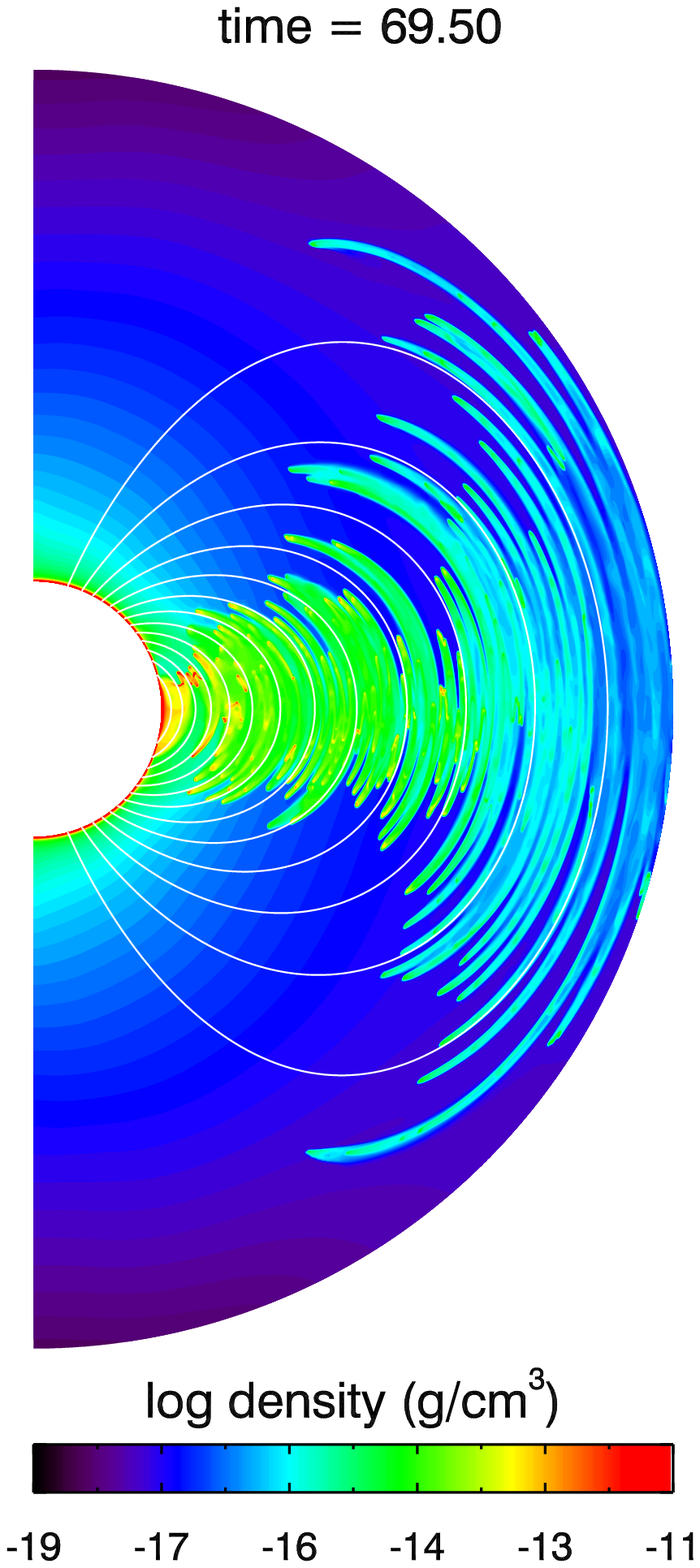}
    \includegraphics[width=.23\textwidth]{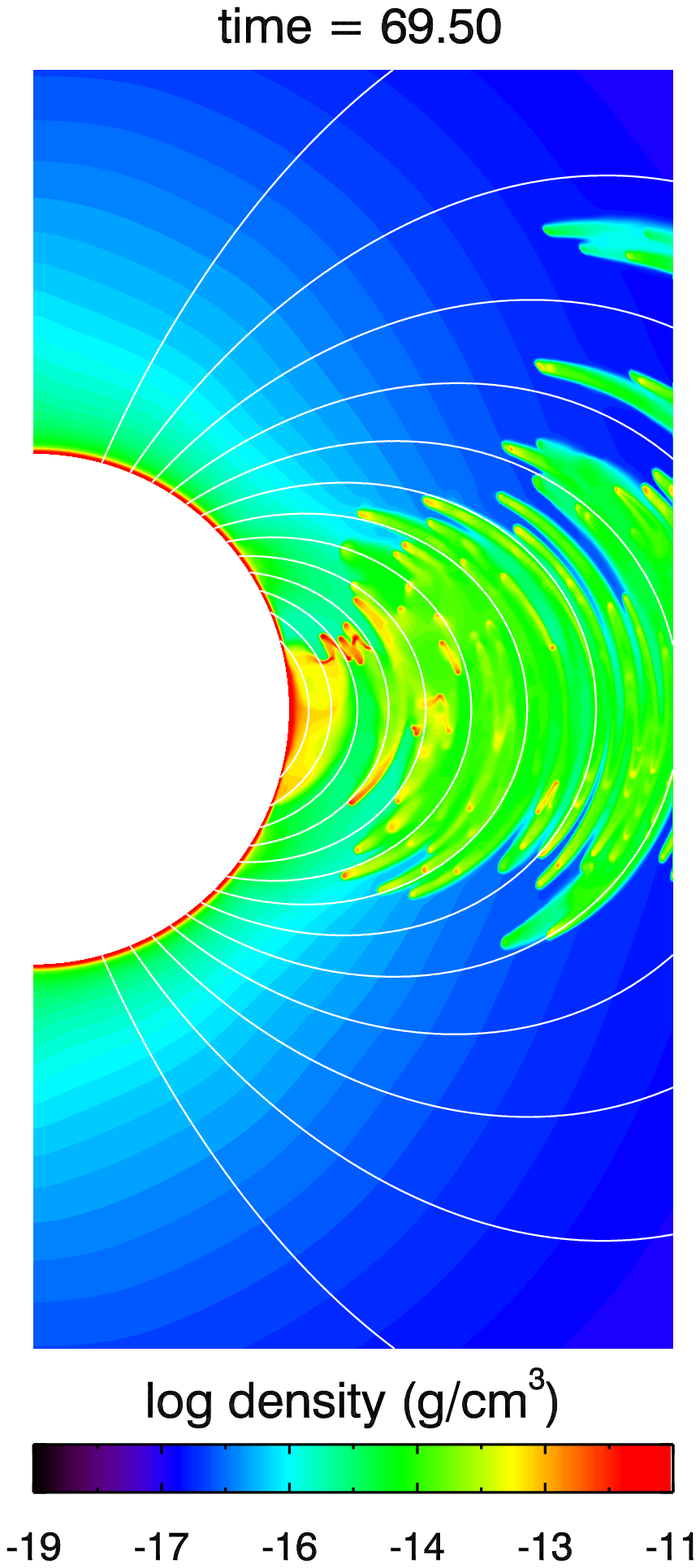}
    \caption{
	Mass density and magnetic field structure in the vicinity of the 
	star. {\em Left}: the whole simulation box. {\em Right}: zoom into the central 
	region. The unit of the time stamps is ks.
		}
    \label{fig:sim1}
\end{figure}

In the inital phase of the simulation run, the gas flow and mass density 
undergo drastic changes. While the gas density at high latitude drops, gas 
piles up in the equatorial plane, where a thin disk forms from which 
filaments grow  that are aligned with the magnetic field. 
Figure~\ref{fig:sim1} shows a snapshot of the mass density taken at a time  
when the system has reached a semi-stationary state insofar as the general 
mass distribution with a high mass concentration at low to mid latitudes does 
not evolve much any more. The shape and length of the individual filaments 
is highly variable, though, and the system does not reach a truly stationary 
state. Note that all the field lines shown are closed. Compared to the original 
dipole, the field is slightly compressed by the gas load but the general 
field topology is unchanged. This is generally different for weak surface 
fields or at greater distances from the star, where the field lines are 
stretched away from the star and a thin disk of outflowing material forms.

\begin{figure}
\centering
   \includegraphics[width=.23\textwidth]{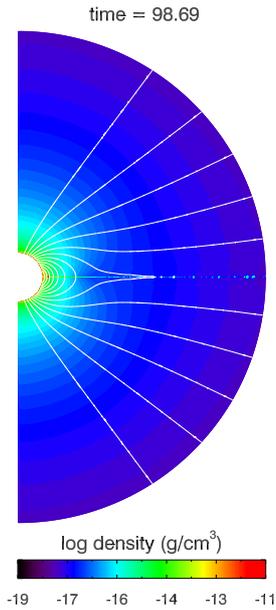}
   \caption{
	Mass density and magnetic field from a simulation with 
	$B_0 = {\rm 1kG}$.
		}
   \label{fig:sim2}
\end{figure}

Figure~\ref{fig:sim2} shows a snapshot from a simulation with a weaker field of 
1\,kG. The magnetosphere is much less extended both in radius and latitude 
and far away from the star the magnetic field is stretched out into a 
configuration resembling a split monopole with a current sheet in the 
equatorial plane. The gas forms a thin disk of enhanced mass density in the 
equatorial plane. The disk is fragmented and the magnetic field is frozen 
into the fragments. In the equatorial plane the magnetic field forms an 
elongated tail. At the tip, reconnection occurs and magnetic flux is carried 
away by the disk material. We expect a similar picture in case of stronger 
fields, but at distances from the star outside our current simulation box.
Note also that for the stronger magnetic field in Fig.~\ref{fig:sim1}, the 
regions of the stellar surface where open field lines originate is limited to 
a small range of latitudes around the poles. We expect this effect to be even 
stronger with a dipole field strength as large as 20\,kG. The right panel of 
Fig.~\ref{fig:sim1} shows a zoom-in to the inner part of the simulation box.

At the time shown, the magnetosphere is still forming and the filaments 
become longer. Close to the star the mass distribution is quite complex, 
with an increased mass density at low latitudes close to the stellar surface 
and blobs of gas at low to mid-latitudes. Within the filament structure, gas 
moves back and forth along the field lines and inward towards the star. The 
mass distribution is distinctly asymmetric with respect to the equatorial 
plane. At this point we can not predict the final structure and extent of the 
magnetosphere but we notice the high mass density at low to mid-latitudes.
To summarize, while there must be strong currents present to exert a Lorentz 
force strong enough to trap the gas, we do not find any changes of the field 
topology close to the star that would hint at possible reconnection events in 
that region. 


\section{Discussion}
\label{sec:disc}

Slow rotators with strong magnetic fields, such as \mbox{NGC\,1624-2}, form 
dynamical magnetospheres, in which material flows along closed magnetic 
field loops from both magnetic poles, colliding near the magnetic equator and 
then falling back onto the stellar surface, leading to complex flows in the 
magnetosphere. \mbox{NGC\,1624-2}'s magnetosphere was estimated to extend to 
an Alfv\'en radius $R_{\rm A}$ of more than eleven stellar radii, hence 
trapping 95\% of the outflowing wind, much more than other magnetic O stars 
that have an $R_{\rm A}$ of just a few stellar radii 
\citep{David-Uraz2019}. 
A much larger and denser magnetosphere compared to that of any other magnetic 
O-type star was also confirmed by Chandra observations, where the high X-ray 
luminosity, its variation with stellar rotation, and its large attenuation are 
consistent with a large dynamical magnetosphere with magnetically confined 
wind shocks
\citep{Petit2015}.

\mbox{NGC\,1624-2} is the only magnetic O-type star for which the variability
of both the longitudinal magnetic field and the mean magnetic field modulus 
is currently studied. The approximately sinusoidal variation of $\bz$ and 
the ratio of the values of the extrema of $\langle B \rangle$ indicate that 
there is an important component of the field that is dipolar. The absence of 
a sign reversal of the mean longitudinal field over the rotation cycle 
indicates that only one magnetic pole is visible over the rotation cycle. 
Involving the mean field modulus measurements, we find that the geometrical 
structure of the magnetic field of \mbox{NGC\,1624-2} must depart from a 
centred dipole. A similar conclusion has been independently obtained by
\citet{daviduraz2020}.
Notably, among the eleven magnetic O-type stars with known $\bz$ variation 
curves,for seven stars, HD\,108, HD\,37022 (=$\theta^1$\,Ori\,C), HD\,54879, 
HD\,148937, HD\,191612, Tr\,16-22, and \mbox{NGC\,1624-2}, only one magnetic 
pole is visible throughout the rotation cycle
\citep[][this work]{ShultzWade2017,Petit2008,Hubrig2020,Wade2012b,Donati2006,Naze2016}.
This implies that, due to an unfavorable inclination of the rotation axis,
a larger fraction of their surfaces can never be observed, leaving the 
structure of the field over the invisible surface only constrained by the 
assumption of a tilted centred dipole. There will thus always be a 
considerable degree of ambiguity left in the magnetic models of these stars.

Our $\bz$ values measured using \ion{He}{i} emission lines are predominantly 
negative and, similar to the measurements of other spectral lines, show clear 
variability over the rotation period, with the negative $\bz$ extremum 
observed around phase 0, where the photospheric absorption lines show a 
positive $\bz$ extremum. To explain the presence of variable Stokes~$V$ 
signatures corresponding to a longitudinal field of negative polarity in 
these lines, we suggest three alternative scenarios: one is related to a 
composite structure representing the contribution of several components, 
another one referring to the complex structure of the magnetospheric 
circumstellar environment of \mbox{NGC\,1624-2} with the possible presence of 
a reconnection-driven emission process, and a third one where the \ion{He}{i} 
emission comes from circumstellar material accumulated in the equatorial 
region of the magnetosphere, where the dipolar magnetic field shows an 
inverted sign with respect to the pole.

As of today, magnetic reconnection is best studied in the solar photosphere, 
but it is not clear how the scenarios developed for the Sun can apply for 
massive stars. Although the detection of a negative longitudinal field 
inferred from the \ion{He}{i} emission lines was reported about eight years 
ago, neither magnetohydrodynamical (MHD) simulations nor analytical models 
have been developed to provide a description of the magnetospheric structure 
permitting the reproduction of the observed phenomenon. According to 
\citet{Owocki2016} 
the presence of a huge magnetosphere and a very strong surface field are 
prohibitively expensive to model using numerical MHD. Our MHD numerical 
simulations show that while reconnection events indeed occur in 
magnetospheres of massive stars and the magnetic flux is carried away by the 
disk material, such events take place too far from the stellar surface where 
\ion{He}{i} emission lines are formed.

On the other hand, as the shape of the Stokes~$V$ profiles of the \ion{He}{i} 
emission lines displayed in Figs.~\ref{afig:IVN5875}--\ref{afig:IVN7065} 
cannot be explained in terms of the Zeeman effect, the alternative hypothesis 
related to a composite structure of Stokes~$I$ and $V$ profiles presenting 
the contribution of several components appears currently more promising.
It is possible that future observations of all four Stokes parameters in 
these lines will help to disentangle the contribution of different line 
components.


\section*{Acknowledgements}

We would like to thank the anonymous referee for their  rigid review of our 
article, especially for their suggestions concerning the interpretation of the 
inverted magnetic field in the helium lines.
We also thank G.~Mathys for a discussion on the field modulus.
SPJ is supported by the German Leibniz-Gemeinschaft, project number P67-2018.
Based on observations collected at the Canada-France-Hawaii Telescope (CFHT),
which is operated by the National Research Council of Canada, the Institut
National des Sciences de l'Univers of the Centre National de la Recherche
Scientifique of France, and the University of Hawaii.
Also based on archival observations obtained at the Bernard Lyot Telescope 
(Pic du Midi, France) of the Midi-Pyr\'en\'ees Observatory, which is operated 
by the Institut National des Sciences de l'Univers of the Centre National de 
la Recherche Scientifique of France.
Based on data acquired with the Potsdam Echelle Polarimetric and 
Spectroscopic Instrument (PEPSI) using the Large Binocular Telescope (LBT) in 
Arizona.
This work has made use of the VALD database, operated at Uppsala University,
the Institute of Astronomy RAS in Moscow, and the University of Vienna.


\section*{Data Availability}

The ESPaDOnS data underlying this article are available in the CFHT Science 
Archive at https://www.cadc-ccda.hia-iha.nrc-cnrc.gc.ca/en/cfht/ and can be 
accessed with the object name. Similarly, the Narval data as well as
the ESPaDOnS are available in PolarBase at http://polarbase.irap.omp.eu. The 
PEPSI and ARCES data underlying this article will be shared on a reasonable 
request to the corresponding author.

\appendix
\section{Profiles for longitudinal magnetic field measurements}

\begin{figure*}
 \centering 
 \includegraphics[width=\textwidth]{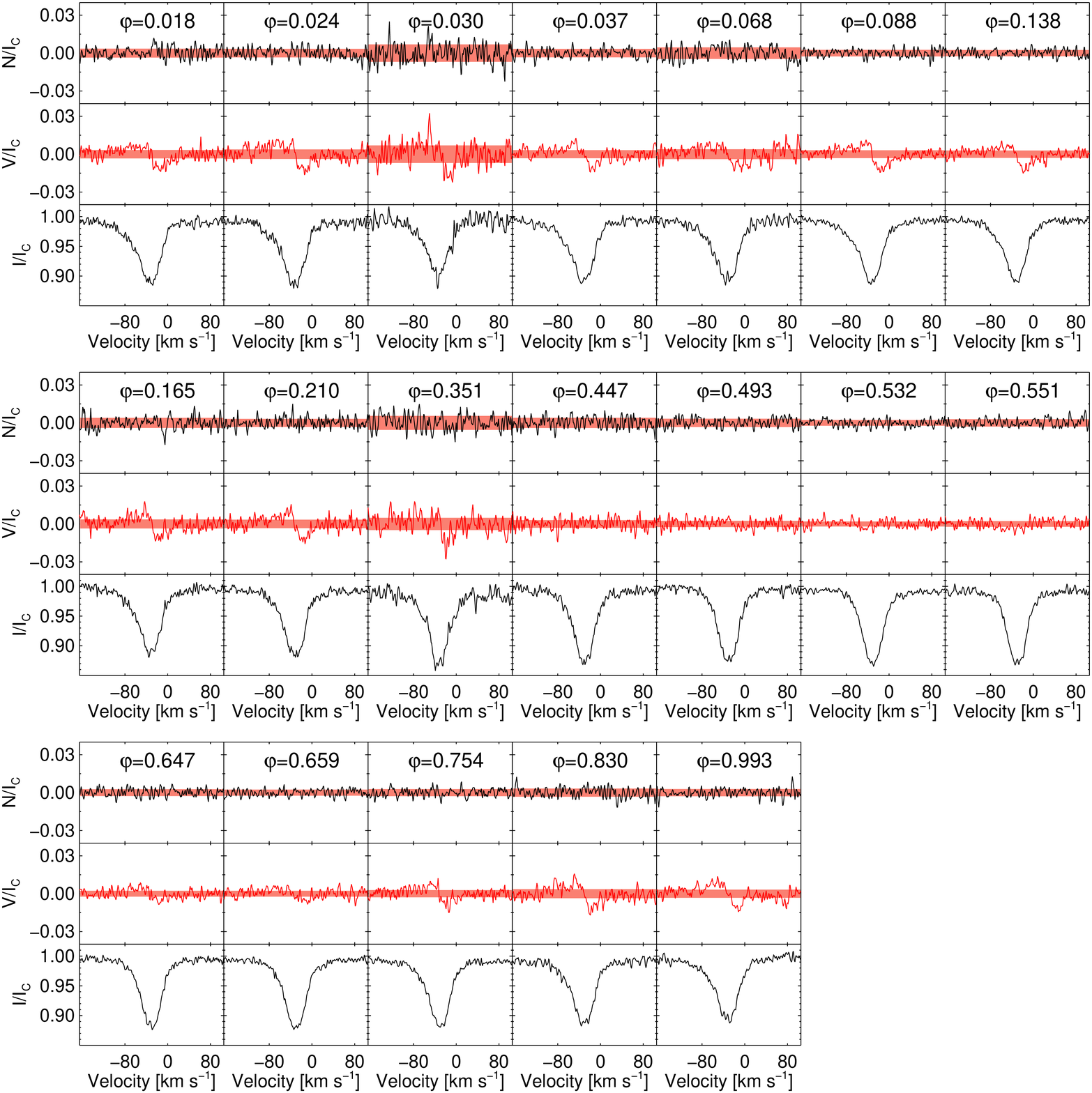}
        \caption{LSD Stokes~$I$, $V$, and null profiles for the line list with 
absorption lines. The plots are sorted by rotational phase. Shaded 
regions in Stokes~$V$ and null panels indicate the mean uncertainty.
                  }
   \label{afig:IVNabsl}
\end{figure*}

\begin{figure*}
 \centering 
 \includegraphics[width=\textwidth]{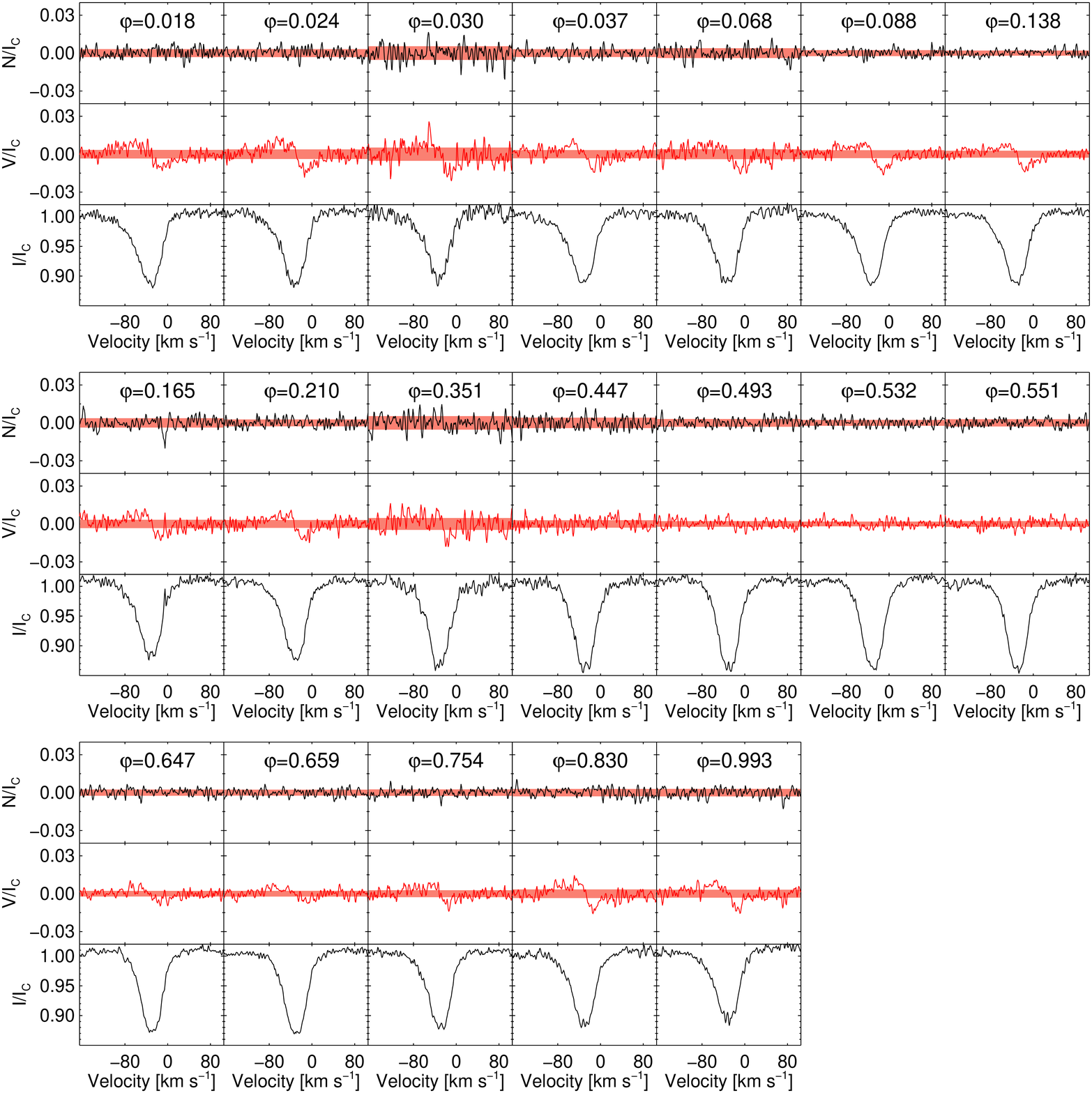}
        \caption{As Fig.~\ref{afig:IVNabsl}, but for the line list without
Si absorption lines.
                  }
   \label{afig:IVNabss}
\end{figure*}

\begin{figure*}
 \centering 
 \includegraphics[width=\textwidth]{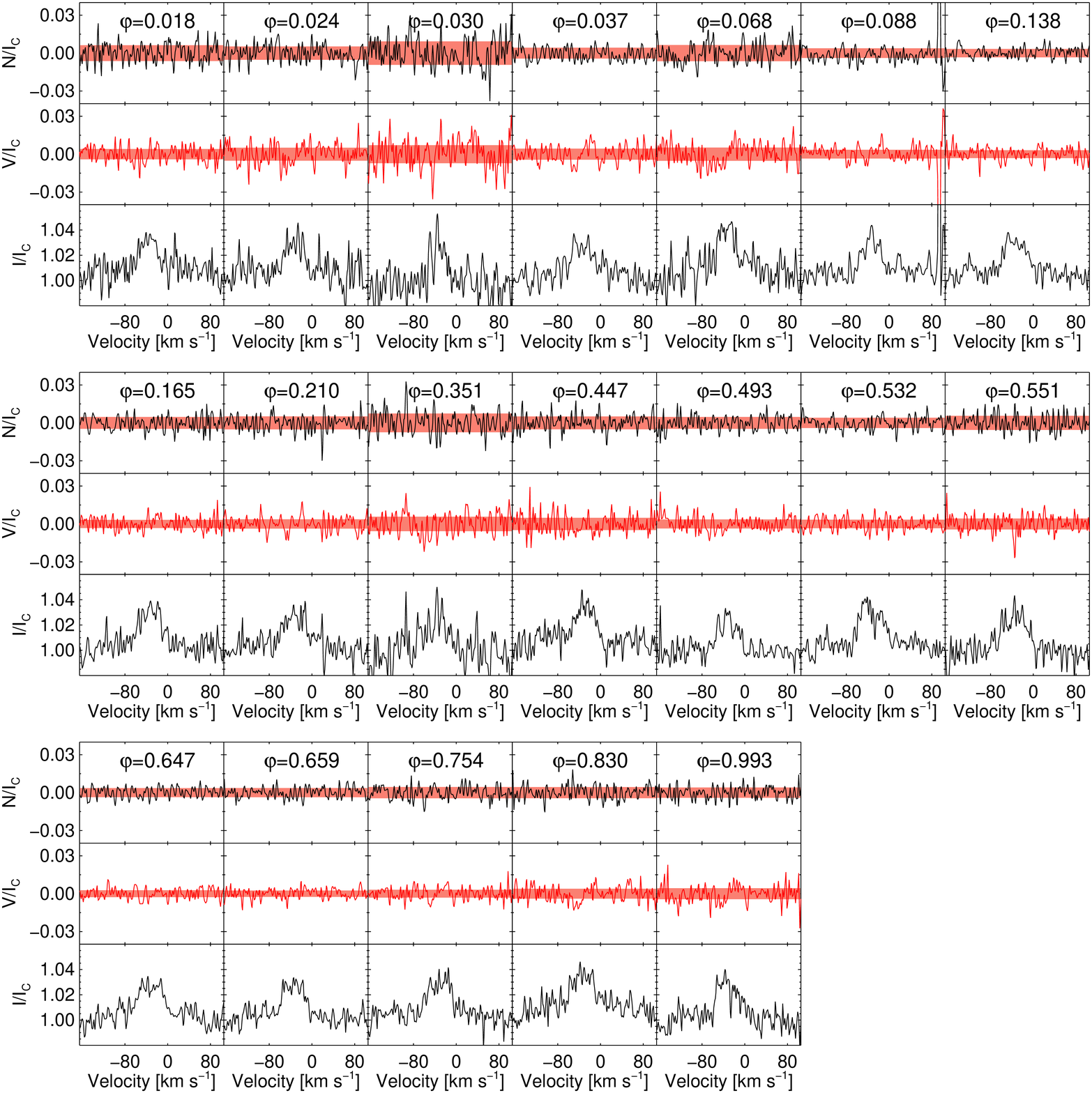}
        \caption{
Stokes~$I$, $V$, and null profiles in velocity frame for the high-excitation 
\ion{O}{iii} $\lambda$7455 emission line (see Fig.~\ref{afig:IVNabsl}).
                  }
   \label{afig:IVNO}
\end{figure*}

\begin{figure*}
 \centering 
 \includegraphics[width=\textwidth]{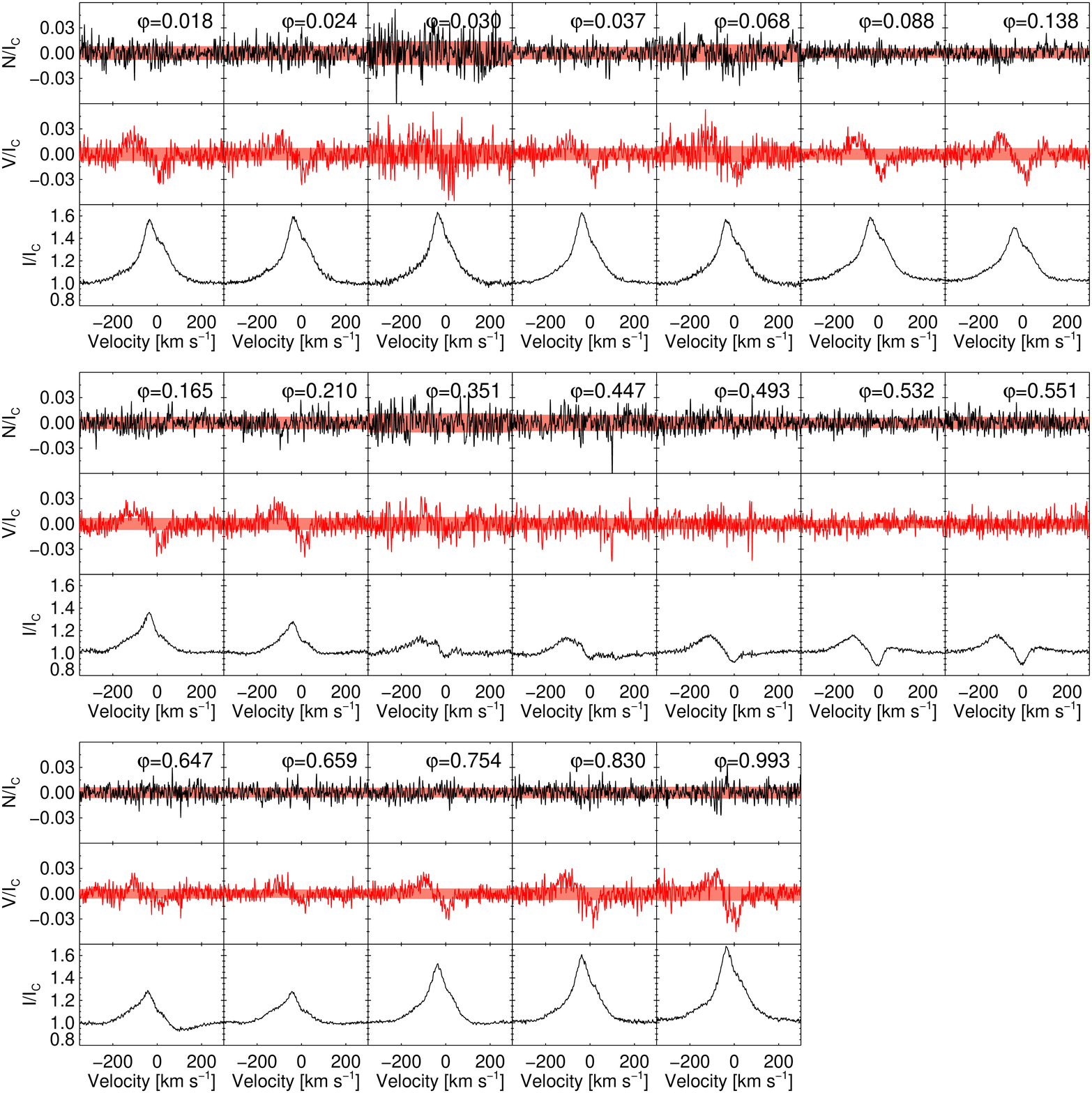}
        \caption{As Fig.~\ref{afig:IVNO}, but for the  
\ion{He}{i} $\lambda$5876 emission line.
                  }
   \label{afig:IVN5875}
\end{figure*}

\begin{figure*}
 \centering 
 \includegraphics[width=\textwidth]{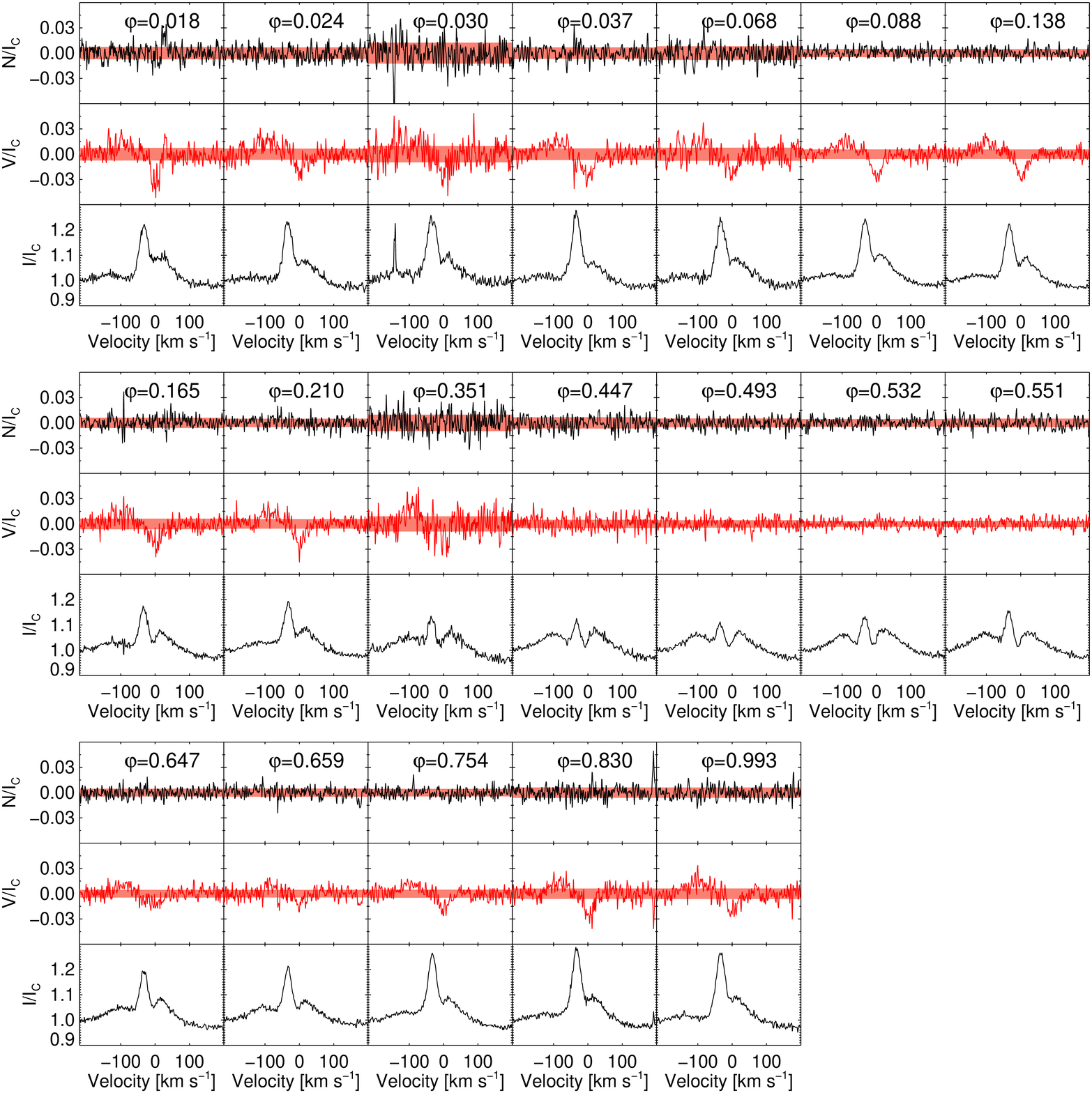}
        \caption{As Fig.~\ref{afig:IVNO}, but for the 
\ion{He}{i} $\lambda$6678 emission line.
                  }
   \label{afig:IVN6678}
\end{figure*}

\begin{figure*}
 \centering 
 \includegraphics[width=\textwidth]{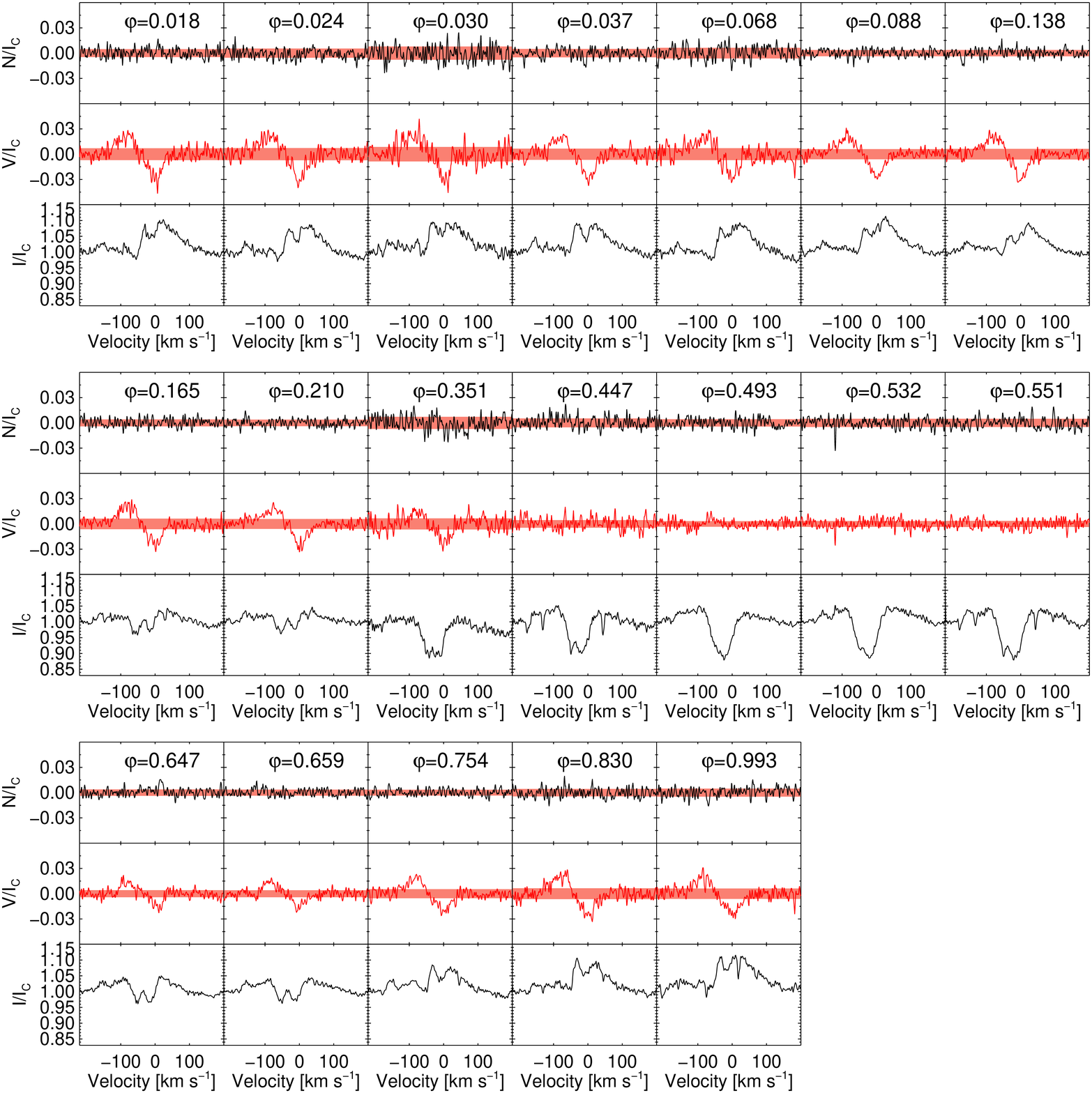}
        \caption{As Fig.~\ref{afig:IVNO}, but for the 
\ion{He}{i} $\lambda$7065 emission line.
                  }
   \label{afig:IVN7065}
\end{figure*}

LSD Stokes~$I$, $V$, and null profiles are presented
in Fig.~\ref{afig:IVNabsl} for the line list with absorption lines,
in Fig.~\ref{afig:IVNabss} for the same line list, but without the Si absorption lines,
in Fig.~\ref{afig:IVNO} for the high-excitation \ion{O}{iii} $\lambda$7455 emission line,
in Fig.~\ref{afig:IVN5875} for the \ion{He}{i} $\lambda$5876 emission line,
in Fig.~\ref{afig:IVN6678} for the \ion{He}{i} $\lambda$6678 emission line, and
in Fig.~\ref{afig:IVN7065} for the \ion{He}{i} $\lambda$7065 emission line.
The plots are sorted by rotational phase.
Shaded regions in Stokes~$V$ and null panels
indicate the mean uncertainty.

\section{Nebular lines}

\begin{figure*}
 \centering 
 \includegraphics[width=\textwidth]{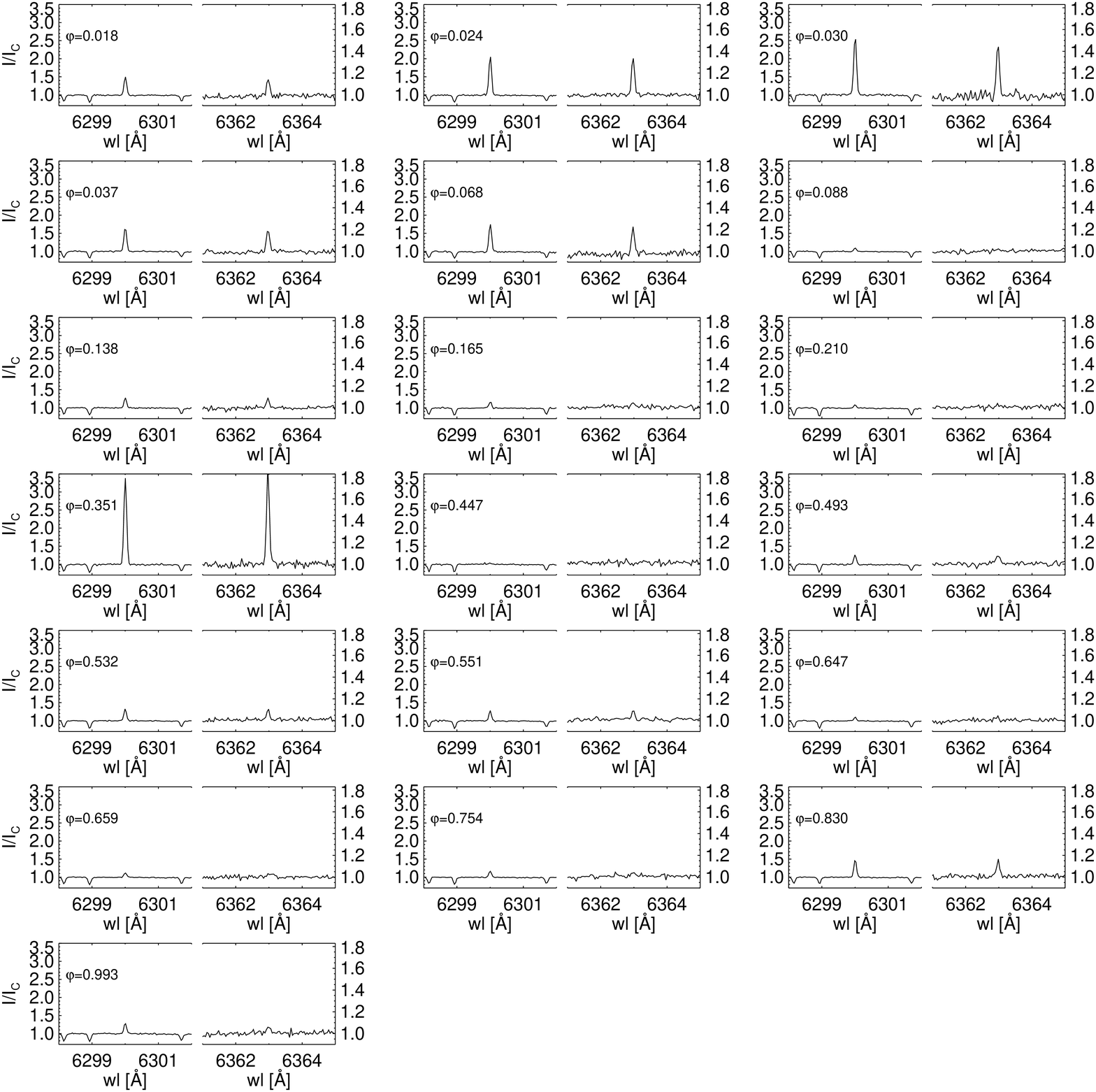}
        \caption{The strength of the forbidden [\ion{O}{i}] nebular lines at 
$\lambda$6300 (left) and $\lambda$6363 (right) as a function of rotation phase.
                  }
   \label{bfig:nebular}
\end{figure*}

The strength of the forbidden [\ion{O}{i}] nebular lines at 
$\lambda$6300 and $\lambda$6363 as a function of rotation phase
is displayed in Fig.~\ref{bfig:nebular}.

\bsp	
\label{lastpage}
\end{document}